\documentclass[11pt]{article}
\usepackage{amsmath, amsthm, amssymb, epsfig, color,latexsym, yhmath}
\allowdisplaybreaks
\newtheorem{prop}{Proposition}[section]

\newcommand{\figref}[1]{Figure~\ref{fig:#1}}

\newcommand{\seclbl}[1]{\label{sec:#1}}

\newcommand{\nno}{\nonumber}
\newcommand{\impls}{\Rightarrow}
\newcommand{\bq}{\begin{eqnarray}}
\newcommand{\eq}{\end{eqnarray}}
\newcommand{\bqa}{\begin{eqnarray}}
\newcommand{\eqa}{\end{eqnarray}}

\newcommand{\tends}{\rightarrow}
\newcommand{\ben}{\begin{enumerate}}
\newcommand{\een}{\end{enumerate}}
\newcommand{\bei}{\begin{itemize}}
\newcommand{\eei}{\end{itemize}}

\newtheorem{thm}{Theorem}[section]

\newcommand{\ie}{i.e., }

\newcommand{\rref}[1]{(\ref{#1})}

\newcommand{\lbr}{\left \{ }
\newcommand{\rbr}{\right \} }
\newcommand{\Lbr}{\left [ }
\newcommand{\Rbr}{\right ] }
\newcommand{\lp}{\left (}
\newcommand{\rp}{\right )}

\renewcommand{\th}{^{{\rm th}}}
\newcommand{\CN}{\mathcal{CN}}
\newcommand{\nmin}{\min\lp n_r, n_t \rp}
\newcommand{\eps}{\epsilon}
\newcommand{\dotgq}{\dot{\geq}}
\newcommand{\dotlq}{\dot{\leq}}

\newcommand{\bw}{{\bf w}}
\newcommand{\bx}{{\bf x}}
\newcommand{\bu}{{\bf u}}
\newcommand{\bone}{{\mathbb{I}}}
\newcommand{\bh}{{\bf h}}
\newcommand{\bd}{{\bf d}}

\newcommand{\by}{{\bf y}}
\newcommand{\zQ}{{\bf Q}_Z}
\newcommand{\zP}{{\bf P}_Z}
\newcommand{\vx}{{\bf x}}
\newcommand{\vw}{{\bf w}}
\newcommand{\bp}{{\bf p}}
\newcommand{\mPsi}{{\bf \Psi}}
\newcommand{\mPhi}{{\bf \Phi}}
\newcommand{\mQ}{{\bf Q}}
\newcommand{\mU}{{\bf U}}
\newcommand{\mT}{{\bf T}}
\newcommand{\mM}{{\bf M}}

\newcommand{\mV}{{\bf V}}
\newcommand{\mLambda}{\mathbf{\Lambda}}

\newcommand{\mX}{{\bf X}}
\newcommand{\mR}{{\bf R}}

\newcommand{\mA}{{\bf A}}

\newcommand{\mH}{{\bf H}}

\newcommand{\mD}{{\bf D}}
\newcommand{\mI}{{\bf I}}

\newcommand{\Co}{{\bf C}}

\newcommand{\beqan}{\begin{eqnarray*}}
\newcommand{\eeqan}{\end{eqnarray*}}
\newcommand{\beqa}{\begin{eqnarray}}
\newcommand{\eeqa}{\end{eqnarray}}
\newcommand{\bear}{\begin{eqnarray}}
\newcommand{\ear}{\end{eqnarray}}
\newcommand{\bears}{\begin{eqnarray*}}
\newcommand{\ears}{\end{eqnarray*}}
\newcommand{\beq}{\begin{equation}}
\newcommand{\eeq}{\end{equation}}

\newcommand{\dmin}{d_{\min}}

\newcommand{\CH}{{\mathcal H}}

\newcommand{\vpsi}{\mbox{\boldmath$\psi$}}
\newcommand{\vphi}{\mbox{\boldmath$\phi$}}
\newcommand{\vlambda}{\mbox{\boldmath$\lambda$}}
\newcommand{\valpha}{\mbox{\boldmath$\alpha$}}

\newcommand{\vy}{{\bf y}}

\newcommand{\vh}{{\bf h}}

\newcommand{\vd}{{\bf d}}

\newcommand{\vq}{{\bf q}}

\newcommand{\SNR} {{\sf SNR}}

\newcommand{\E}{\Bbb{E}}
\newcommand{\prob}[1]{\mathbb{P} \left\{ #1 \right\}}

\newcommand{\df}{:=}

\title{Approximately Universal Codes over Slow Fading Channels}

\author{Saurabha Tavildar and Pramod Viswanath
\thanks{The authors are with the department of Electrical and Computer Engineering, and
 the Coordinated Science Laboratory at the University of Illinois at
 Urbana-Champaign. Email: {\tt \{tavildar, pramodv\}@uiuc.edu}. The
 material in this paper has appeared at the {\em Conference on
 Information Sciences and Systems}, Princeton 2004. Chapter~9 of a
 recent book, {\em Fundamentals of Wireless Communication},
 Cambridge University Press 2005, is based in part on this work.
 This research was supported in part by the National Science
 Foundation under grants NSF CAREER 0237549 and NSF ITR 0325924,  by
 a Vodafone graduate fellowship, and by  Motorola Inc.}}

\begin{document}

\maketitle

\begin{abstract}
Performance of reliable communication over a coherent slow fading
MIMO channel at high SNR is succinctly captured as a fundamental
tradeoff between diversity and multiplexing  gains. We study the
problem of designing codes that optimally tradeoff the  diversity
and multiplexing gains. Our main contribution is a precise
characterization of codes that are {\em universally}
tradeoff-optimal, \ie they optimally tradeoff the diversity and
multiplexing gains for {\em every} statistical characterization of
the fading channel. We denote this characterization as one of {\em
approximate universality} where the approximation is in the
connection between error probability and outage capacity with
diversity and multiplexing gains, respectively. The characterization
of approximate universality is then used to construct new coding
schemes as well as to show optimality of  several schemes proposed
in the space-time coding  literature.
\end{abstract}

\section{Introduction}

Reliable communication over {\em slow} fading point-to-point channels,
where the (random) channel realization is fixed over the time scale
of communication, is characterized by the tradeoff between data rate
and error probability:  typical fading
distributions have a nonzero probability of being very small and
thus arbitrarily reliable communication is not possible at any
non-zero rate. The tradeoff between the data rate and the error
probability is captured by the outage capacity, the largest rate of
reliable communication for a fixed error probability. The
information theoretic view is that of a {\em compound} channel: the slow
fading channel is composed of a class of channels parameterized by
the different channel realizations that are not in outage. The
outage capacity is achieved by {\em universal} codes, those that
work reliably over {\em every} one of the channel realizations not
in outage.

At high SNR, the precise (but too involved to derive code design
principles) tradeoff between error probability and data rate is
coarsely captured in terms of a tradeoff between diversity and
multiplexing gains \cite{ZT03}: these are the rate of decay of error
probability and the increase of data rate with increasing SNR. Since
the tradeoff is captured at a coarser scale, we shall denote codes
that optimally tradeoff diversity and multiplexing gains for {\em
every} slow fading channel as {\em approximately universal}; the
approximation here refers to the coarseness in the definition of
diversity and multiplexing gains as opposed to studying error
probability and data rate directly. Our main result is a {\em
precise} characterization of approximately universal codes. We use
this characterization to show the approximate universality of some
codes proposed in the literature and to also construct new
space-time codes that are approximately universal. These codes are
robust to statistical channel modeling errors, hence their
engineering appeal is clear. This approach of using compound channel
viewpoint to construct robust codes for MIMO channels has  also been
taken
%by Wesel et al.
in a series of works in \cite{W96, KW03, MW03}.

We are interested in codes that achieve reliable communication over
all channel realizations not in outage: this suggests, as done in
\cite{KW03}, asking for the performance of the code for the {\em
worst} channel not in outage. This is in contrast to the traditional
performance analysis where the error probability is {\em averaged}
over the statistics of the fading channel. In particular, if the
worst-case pairwise error probability decays {\em exponentially}
with increasing SNR then such a code is approximately universal. For
a parallel channel, the worst channel for a given pair of codewords
is ``inverse waterfilling'' over the pairwise squared codeword
differences. For a MIMO channel, the worst channel (derived in
\cite{KW03}) aligns its singular vectors in the same directions as
those of the pairwise codeword difference matrix and then the
singular values inverse waterfill the singular values of the
pairwise codeword difference matrix. While the exact expression of
the worst-case pairwise error is somewhat involved, a simple
worst-case code design criterion emerges at high SNR for both the
parallel channel and the MIMO channel.

For a parallel channel, somewhat surprisingly, the worst-case code
design criterion at high SNR simplifies to the product distance
criterion which was derived initially for the i.i.d.\ Ricean fading
channel \cite{DS88}, though is better known for the i.i.d.\ Rayleigh
fading channel (see Chapter~3 of \cite{TV05}). In a compound channel
setting the criterion was heuristically derived in \cite{W96}, here
we give a more precise statement for the criterion. In particular,
we show that if the products of all normalized squared codeword
differences is larger than $2^{-R}$ where $R$ is the communication
rate, then the code is approximately universal. This design
criterion suggests a class of codes based on permutations of the QAM
(quadrature amplitude modulation) constellation that we call {\em
permutation codes}. Even random permutation codes are approximately
universal and we provide examples of simple and explicit permutation
codes that are approximately universal. We show that a code based on
a rotated QAM constellation proposed in the literature \cite{YW03}
also satisfies the desired product distance property and is hence
approximately universal.

For a MIMO channel, the worst-case code design criterion is in
general not simply to maximize the determinant of the codeword
difference matrix, the criterion derived for the i.i.d.\ Rayleigh
fading channel \cite{TCS98}. This can be explicitly seen in the case
of the multiple transmit but single receive antenna (MISO) channel:
the worst channel chooses the most susceptible direction to confuse
between a pair of codeword matrices -- this is the direction of the
smallest singular value of the codeword difference matrix. Thus the
worst-case code design criterion for the MISO channel is to maximize
the {\em smallest} singular value of the codeword difference matrix;
different  from the determinant criterion derived for the i.i.d.\
Rayleigh fading channel. More generally, the worst-case code design
criterion at high SNR for a MIMO channel (with $n_t$ transmit and
$n_r$ receive antennas) is to maximize the product of the smallest
$\min(n_t,n_r)$ singular values of the codeword difference matrix.
With more receive than transmit antennas, the worst-case code design
criterion reduces to the determinant criterion derived for the
i.i.d.\ Rayleigh fading channel.

An important implication of our worst-case code design criterion is
the following: if a code is  approximately universal on an
$n_t\times n_t$ MIMO channel, then it is also approximately
universal for $n_t\times n_r$ MIMO channel for {\em every} $n_r$.
Several space-time codes proposed in the literature satisfy the
worst-case code design criterion and are hence approximately
universal. In particular, the QAM rotation codes in \cite{YW03,DV03}
are approximately universal for every MIMO channel with two transmit
antennas. The recently proposed codes in  \cite{EKPKL04, ORBV04,
KR05} that are derived from cyclic division algebra are also
approximately universal.

\begin{figure}[h]
\begin{center}
\scalebox{0.8}{\input{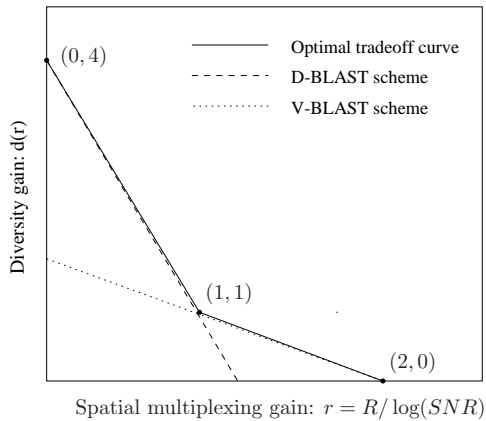}} \caption{Tradeoff
curves: $n_t = n_r = 2$} \label{fig:ntnr22}
\end{center}
\end{figure}

V-BLAST \cite{FGRW99} and D-BLAST \cite{F96} are classical
architectures for communication over a MIMO channel. While they are
not approximately universal, we show that they are tradeoff optimal
in some rate regime universally over a (restricted) class of
channels which are rotationally invariant. In particular, this class
of channels includes the i.i.d.\ Rayleigh fading channel: we show
that V-BLAST with simple QAM constellations as the independent data
streams achieves the last segment of the tradeoff curve for the $n
\times n$ i.i.d.\ Rayleigh fading MIMO channel  and D-BLAST achieves
the first segment of every $n_t\times 2$ i.i.d.\ Rayleigh fading
MIMO channel. These results are illustrated in the context of a
$2\times 2$ i.i.d.\ Rayleigh fading  MIMO channel in
\figref{ntnr22}.

We have organized  this paper into two distinct parts: first, we present a
precise characterization of approximate universal codes for the  general
MIMO channel. In the second part,  we discuss explicit approximately
universal codes,   starting with simpler channel
models and moving on to the more involved ones. In particular, we
start with the scalar channel and show that a simple QAM is
approximately universal (this is done in Section~\ref{sec:scalar}).
Next, we study the parallel channel and the MISO channel in
Sections~\ref{sec:parallel-channel} and~\ref{sec:miso},
respectively. Finally we consider the general MIMO channel in
Section~\ref{sec:mimo} by demonstrating the approximately universality
of some codes proposed in the literature, and then analyzing the
approximately universal performance
of V-BLAST and D-BLAST in Sections~\ref{sec:vblast}
and~\ref{sec:dblast}, respectively.

\section{Channel Model and The Outage Formulation}
The main focus of this paper is on the slow fading (point-to-point)
MIMO channel
\begin{eqnarray}\label{eq:general-cm}
\by[m] = \mH \bx[m] + \bw[m],
\end{eqnarray}
where $m$ is the time index and $\by$ and $\bx$ denote the output
and the input vectors respectively. The complex $n_r \times n_t$
matrix $\mH$ of fading gains is randomly picked, but stays constant
over the time-scale of communication; we suppose that the exact
realization of $\mH$ is known at the receiver. The additive noise
$\bw$ has i.i.d.\ complex Gaussian ($\CN(0,1)$) entries. We are
interested in one-shot communication over this channel over a
(small) length of time $T$. There is a transmit power constraint of
$Tn_t\SNR$ for any transmit codeword of length $T$.

In this paper, we focus on the scaling at high $\SNR$ introduced in
\cite{ZT03}: the data rate is measured on a scale of $\log\SNR$ and
the decay rate of error probability is measured on a scale which is
a negative exponent of $\SNR$. All logarithms in this paper are to
the base 2. More precisely, the {\em multiplexing} and {\em
diversity} gains are defined as follows. A sequence of coding
schemes (sequence in $\SNR$) achieves a multiplexing rate of $r$ and
diversity gain of $d$ if
\begin{eqnarray}
\nno \lim_{\SNR \tends \infty} \frac{R(\SNR)}{\log{\SNR}} =  r, \quad
\mbox{and} \quad
\lim_{\SNR \tends \infty}
\frac{\log{\mathbb{P}_e(\SNR)}}{\log{\SNR}}  =  -d,
\end{eqnarray}
where $R(\SNR)$ is the rate of the scheme and $\mathbb{P}_e(\SNR)$
is the probability of error with maximum likelihood (ML) decoding
for the scheme. For a given multiplexing gain $r$, the largest
diversity gain supported by any coding scheme is denoted by
$d^*(r)$. The goal is to find a characterization of this optimal
diversity-multiplexing tradeoff, $d^*(r)$, for any correlated
channel and then to find (simple) coding schemes with as small a
block length ($T$) as possible that achieve this optimal tradeoff
curve.

The outage event turns out to be closely related to the problem of
characterizing $d^*(r)$. It is defined as the set of channel
realizations for which the mutual information  is below the data
rate:
\begin{equation}\label{eq:outage}
\lbr \mH  : I\lp \bx ; \mH\bx + \bw |\mH  \rp <  R\rbr,
\end{equation}
where the input distribution is independent of the realization of
$\mH$. It is shown in \cite{ZT03} that, in the scale of interest,
the input distribution $\mathbb{P}_{\vx}$ can be taken to be i.i.d.\
complex Gaussian for the Rayleigh fading channel; a similar argument
for any fading distribution shows that the input distribution can be
taken to be i.i.d.\ complex Gaussian. This means that the outage
curve can be defined as: \beq\label{eq:outage_curve_formula} d_{\rm
out}(r) \df \lim_{\SNR \rightarrow \infty} \frac{-\log\prob{\mH :
\log\det\lp \mI + \SNR \mH\mH^*\rp <
 r\log\SNR}}{\log\SNR}. \eeq

The outage curve $d_{\rm out}(r)$ is an upper bound to $d^*(r)$
\cite{ZT03}. On the other hand, the set of channel realizations that
are not in outage constitute a compound channel, the capacity of
which is $r\log\SNR$. The compound channel coding theorem guarantees
the existence of universal codes: codes that achieve reliable
communication over {\em every} MIMO channel realization that is not
in outage.  This means, that by coding over possibly long block
lengths, one can actually achieve the outer bound of $d_{\rm
out}(r)$. Therefore for the rest of this paper, {\em we identify the
outage curve with the optimal diversity-multiplexing tradeoff
curve}. Note that, we are mainly interested in fading distributions
such that the eigen-values are not bounded away from zero (e.g.\
AWGN channel can be considered as a fading channel). Otherwise, the
outage curve will be infinite, and an approximately universal code
will achieve it. But, the diversity-multiplexing tradeoff is not the
right setup to study this problem.

We are interested in  universal codes that achieve  the upper bound
of $d_{\rm out}(r)$ only to the extent that they are
tradeoff-optimal; we call such codes  approximately universal. Our
main focus is on a characterization of approximately codes with {\em
small block-length}.

\section{Main Result}
Our main result is a precise characterization of approximately
universal codes.
\begin{thm}\label{thm:au_mimo}
A sequence of codes of rate $R(\SNR)$ bits/symbol is approximately
universal over the MIMO channel if and only if, for every pair of
codewords, \beq\label{eq:mimo_univ_crit} \lambda_1^2
\lambda_2^2\cdots \lambda_{\nmin}^2  \geq
\frac{1}{2^{R(\SNR)+o(\log(\SNR))}}, \eeq where $\lambda_1,\ldots
,\lambda_{\nmin}$ are the smallest $\nmin$ singular values of the
normalized (by $\frac{1}{\sqrt{\SNR}}$) codeword difference matrix.
\end{thm}

For $n_r \geq n_t$, \rref{eq:mimo_univ_crit} turns out to be the
same as the ``nonvanishing determinant'' criterion introduced in
the context of i.i.d.\ Rayleigh fading channels in \cite{BR03}. This
criterion was also studied in \cite{YW03, EKPKL04}, also in the
context of i.i.d.\  Rayleigh fading channels. In \cite{YW03},
it was shown that for two transmit antennas, if a code satisfies
this nonvanishing determinant criterion, then it is
tradeoff-optimal for the i.i.d.\ Rayleigh fading channel; this
result has been recently generalized for artibtrary number of
transmit antennas in \cite{EKPKL04}.

Our result is much
stronger: if a code satisfies the nonvanishing
determinant criterion, then it is tradeoff-optimal for {\em every}
fading distribution. Thus, our result gives the well-known
determinant criterion a precise {\em operational} interpretation in terms
of approximate universality. Through this characterization, we will
see that codes with small block lengths can be approximately
universal. We start with a few implications of this criterion and
then prove the sufficiency part of the
criterion. The necessity part is proved in Appendix
\ref{ap:precise_char}.

\subsection{Approximately Universal Codes in the
Downlink}\label{sec:dl_universality} Some interesting observations
follow from our characterization of approximately universal codes.
\begin{itemize}
\item If a code is  approximately universal over an $n_t\times
n_r$ MIMO channel with $n_r \geq n_t$, i.e.,  the number of receive
antennas is equal to or larger than the number of transmit antennas,
then it is   also approximately universal for an $n_t \times l$ MIMO
channel with $l \geq n_t$.
\item The singular values of the normalized codeword difference
matrices are upper bounded by a  fixed number ($\sqrt{n_tT}$). Thus,
a code that is approximately universal over an $n_t\times n_r$ MIMO
channel is also approximately universal over an $n_t\times l$ MIMO
channel with  $l\leq n_r$.

\item Consider the downlink of a cellular system where the base
stations are equipped with multiple transmit antennas. Suppose we
want to broadcast common information to all the users in the cell.
We would like our transmission scheme to not depend on the number of
receive antennas at the users: each user could have a different
number of receive antennas, depending on the model, age, and type of
the mobile device.  Universal MIMO codes provide an attractive
solution to this problem. Suppose we broadcast the common
information at rate $R$ using an approximately universal  space time
code over an $n_t\times n_t$ MIMO  channel. Since this code is
approximately universal for every $n_t\times n_r$ MIMO channel, the
diversity seen by each user is {\em simultaneously} the best
possible at rate $R$. To summarize: the diversity gain obtained by
each user is the best possible with respect to both,
\begin{itemize}
\item the number of receive antennas the user has, and
\item the statistics of the  fading channel the user is currently
  experiencing.
\end{itemize}

\end{itemize}

\subsection{Characterization of approximately universal codes}
Towards our goal of characterizing approximately universal codes, we
first calculate the pairwise error probability for a pair of
codewords based on the {\em worst} channel realization not in
outage, \ie we consider the realization (not in outage) as a {\em
function} of the specific pair of codewords so as to yield the worst
pairwise error probability. If this worst-case pairwise error
probability decays exponentially with SNR for every pair of
codewords (we allow the worst channel to change as a function of the
pair of codewords), then a simple union bound argument shows that
the error probability conditioned on the channel realization not in
outage decays exponentially with SNR: the total number of codewords
is only polynomial in SNR; for example if the multiplexing rate is
$r$, the rate is $R = r\log\SNR$ and the total number of codewords
is $\SNR^r$. Since the error probability is lower bounded by the
outage probability, we arrive at a {\em sufficient} condition for
approximate universality of a code:
\begin{quote}
the worst-case (over channels not in outage) pairwise error
probability for every pair of codewords should decay exponentially
with SNR.
\end{quote}
It turns out that this condition is {\em necessary} as well; thus we
have an exact characterization of approximately universal codes.

In Section~\ref{sec:mimo-univ-crit} we derive an expression for the
worst-case pairwise error for a pair of codewords. This derivation
allows us to explicitly characterize approximate universality of a
code in terms of a condition on its pairwise difference codewords.
It is fruitful to contrast our approach with the traditional ``code
design criterion'' for space-time codes in the literature where the
pairwise error probability is {\em averaged} over the channel
statistics. This criterion indeed depends on the specific channel
statistics being considered. This is in stark contrast to the
worst-case analysis we have proposed; the corresponding ``universal
code design criterion'' does not depend on the channel statistics
and characterizes properties of a universal code: the engineering
appeal of the universal code design criterion is natural; modeling
channel statistics is a bit of an ``art'' in practice and it is
useful to have a code that is robust to  a variety of channel
statistics.

The classical code design criterion for the i.i.d.\ Rayleigh fading
 channel is the {\em determinant criterion}; as we will see in
Section~\ref{sec:mimo-univ-crit}, the universal code design
criterion at any specific SNR is quite different from the
determinant criterion. However, it is also somewhat involved and is
not directly suited to verify or to design approximately universal
codes. In Section~\ref{sec:proof_au_mimo} we derive a simplified
condition for approximate universality taking the high SNR scaling
into consideration and this high SNR criterion is indeed very
closely related to the determinant criterion.

\subsubsection{Worst-case Pairwise Error Probability}\label{sec:mimo-univ-crit}
Our approach is to study the worst-case pairwise error probability
of the code over MIMO channel realizations not in outage. The
pairwise error probability between two codeword matrices $\mX_A$ and
$\mX_B$ (of length  $T \geq n_t$), conditioned on a specific
realization of the MIMO channel $\mH$, is \beq\label{eq:prwse_pe}
Q\lp\sqrt{\frac{\SNR}{2}\|\mH\mD\|^2}\rp, \eeq where $\mD$  is the
normalized codeword difference matrix \beq \nonumber \mD  =
\frac{1}{\sqrt{\SNR}}\lp \mX_A - \mX_B\rp.\eeq Expanding the channel
and codeword difference matrices using the singular value
decomposition (SVD), \beq\label{eq:svd} \mH \df \mU_1\mPsi\mV_1^*
\quad {\rm and}\quad \mD \df \mU_2\mLambda\mV_2^*,\eeq the pairwise
error probability in \eqref{eq:prwse_pe} can be rewritten as
\beq\label{eq:prwse_pe_again}
Q\lp\sqrt{\frac{\SNR}{2}\|\mPsi\mV_1^*\mU_2\mLambda\|^2}\rp. \eeq
Suppose the  singular values are increasingly ordered in $\mLambda$
and decreasingly ordered in $\mPsi$:
$$
\mPsi \df {\rm diag}\lbr \psi_1,\ldots
,\psi_{\nmin},0,\ldots,0\rbr,\quad \mbox{and}\quad  \mLambda \df
{\rm diag}\lbr \lambda_1,\ldots ,\lambda_{n_t}\rbr.
$$
Then the worst-case rotation can be determined and it turns out to
be the one that aligns the weaker singular values of the channel
matrix with the stronger singular values of the codeword difference
matrix \cite{KW03}. More precisely, the channel eigen-directions
$\mV_1$ that maximize the pairwise error probability in
\eqref{eq:prwse_pe_again} is \cite{KW03} \beq \mV_1 = \mU_2. \eeq
Now, the no-outage condition is only a condition on the non-zero
$\nmin$ singular values of the fading matrix and is given by:
\beq\label{eq:nooutage_mimo} \sum_{\ell=1}^{\nmin} \log\lp 1 +
\SNR|\psi_\ell|^2\rp \geq R. \eeq Hence the worst-case pairwise
error probability for the MIMO channel reduces to the optimization
problem \beq\label{eq:optproblem_mimo} \min_{\psi_1,\ldots
,\psi_{\nmin}}\; \frac{\SNR}{2} \sum_{\ell=1}^{\nmin} |\psi_\ell|^2
|\lambda_\ell|^2, \eeq subject to the constraint in
\rref{eq:nooutage_mimo}.

If we define $Q_\ell: =\SNR \cdot |\psi_\ell|^2 |\lambda_\ell|^2$,
then the optimization problem can be rewritten as \beq \nonumber
\min_{Q_1\ge 0, \ldots, Q_{\nmin} \ge 0} \frac{1}{2}
\sum_{\ell=1}^{\nmin} Q_\ell \eeq subject to the constraint
\beq\nonumber \sum_{\ell=1}^{\nmin} \log\lp 1 +
\frac{Q_\ell}{|d_\ell|^2}\rp \geq R. \eeq This is the dual of the
problem of minimizing the total power required to support a target
rate $R$ bits/symbol per sub-channel over a parallel Gaussian
channel; the solution is just standard waterfilling, and is given
by \beq\label{eq:worstchannel} Q_\ell:=\SNR\cdot |\psi_\ell|^2
|\lambda_\ell|^2  = \lp\frac{1}{\lambda } - |\lambda_\ell|^2 \rp^+,
\quad \ell = 1,\ldots , \nmin. \eeq Here $\lambda$ is the Lagrange
multiplier chosen such that the channel in \eqref{eq:worstchannel}
satisfies \eqref{eq:nooutage_mimo} with equality. The worst-case
pairwise error probability is \beq\label{eq:wc_pep}
Q\lp\sqrt{\frac{1}{2} \sum_{\ell=1}^{\nmin} \lp \frac{1}{\lambda} -
|\lambda_\ell|^2 \rp^+} \rp, \eeq where $\lambda$ satisfies:
\begin{eqnarray}\label{eq:lambda}\sum_{\ell=1}^{\nmin} \Lbr \log \lp \frac{1}{\lambda |\lambda_\ell|^2} \rp
\Rbr^+ =R.
\end{eqnarray}
For convenience, we denote the argument of the
$Q\lp\sqrt\frac{\lp\cdot\rp}{2}\rp$ function at the worst-case
channel realization as the universal code construction criterion for
the given difference codeword pair. In general, the goal is to
maximize this universal code construction criterion:
\beq\label{eq:univcriterion} \sum_{\ell=1}^{\nmin} \lp
\frac{1}{\lambda} - |\lambda_\ell|^2 \rp^+. \eeq

\subsubsection{A Closer Look at the Universal Criterion} To get a
feel for  the universal criterion in \eqref{eq:univcriterion},
consider the  simple case when codeword difference eigenvalues have
the same magnitude, i.e., $|\lambda_1| = \cdots = |\lambda_{n_t}|$.
Then $\lambda$ can be explicitly calculated: \beq\nonumber
\frac{1}{\lambda}  = 2^{R/\nmin}|\lambda_1|^2. \eeq Thus the
universal criterion is given by
  \beq\nonumber \nmin \lp 2^{R/\nmin} - 1\rp |\lambda_1|^2, \eeq a simple function
of the magnitude of the normalized codeword difference.  To
understand the situation in general, let us suppose without any loss
of generality that  $|\lambda_1| \leq \cdots \leq
|\lambda_{\nmin}|$. Now consider the largest $k$ such that
\beq\label{eq:define_k} |\lambda_k|^{2} \leq 2^{R/k} \,
|\lambda_1\cdots \lambda_k|^{2/k} \leq |\lambda_{k+1}|^{2}, \eeq
with $|\lambda_{\nmin+1}|$ defined as $+\infty$. Then $\lambda$ can
be calculated explicitly: \beq\label{eq:lambda_k} \frac{1}{\lambda}
= 2^{R/k} |\lambda_1\cdots \lambda_k|^{2/k}, \eeq satisfies
\rref{eq:lambda}. Thus the universal code design criterion turns out
to be \beq\label{eq:distance_k} \lp k\lp 2^{R}
|\lambda_1\lambda_2\cdots \lambda_k|^2\rp^{1/k} - \sum_{\ell = 1}^k
|\lambda_\ell|^2\rp, \eeq a combination of the geometric and
arithmetic means of the magnitudes of the $k$ smallest singular
values of normalized codeword differences. While this calculation
sheds some insight into the nature of the universal code design
criterion, it still does not lend itself to designing or verifying
approximately universal codes. Towards making this expression more
amenable to code design, we would like to develop a high SNR
approximation; this is done next.

\subsubsection{Proof of
Theorem \ref{thm:au_mimo}}\label{sec:proof_au_mimo} Our goal here is
to show that for a sequence of codes satisfying
\rref{eq:mimo_univ_crit}, the probability of error has the same
decay rate as that of the outage probability for all fading
distributions. The probability of error can be upper bounded using a
smart union bound (as in \cite{ZT03}):
\begin{eqnarray}\label{eq:sub}
\mathbb{P}_e & \leq & \prob{\mathcal{O}} + \mathbb{P}(\mbox{error,
$\mathcal{O}^c$}).
\end{eqnarray}
Here we have denoted the outage event by $\mathcal{O}$. Similar to
the union bound, the second term can be upper bounded by a sum of
pairwise errors averaged over all channel realizations not in
$\mathcal{O}$. This sum can be further upper bounded by the sum of
the {\em worst-case } (over all channel realizations not in
$\mathcal{O}$) pairwise error probabilities. For the probability of
error to behave like the probability of outage for every fading
distribution, we {\em require} the second term in \eqref{eq:sub} to
decay exponentially in $\SNR$ ($=e^{-\SNR^{\delta}}$ for some
$\delta> 0$). One way to do this is to make {\em every} worst-case
pairwise error decay exponentially in $\SNR$.

Instead of considering a single outage event, we consider a sequence
of outage events $\mathcal{O}_\eps$, parameterized by $\epsilon >
0$: the    channel  realizations not in $\mathcal{O}_\eps$  are
those that are {\em  strictly} inside the no-outage region:
\beq\nonumber \sum_{\ell=1}^{\nmin} \log\lp 1 + |\psi_{\ell}|^2
\SNR\rp\geq R(1+\epsilon). \eeq For a pair of codewords, the
worst-case pairwise error probability is \rref{eq:wc_pep}
\beq\nonumber Q\lp\sqrt{\frac{\sum_{\ell=1}^{\nmin} \lp
\frac{1}{\lambda} - |\lambda_{\ell}|^2\rp^+}{2}}\rp,  \eeq where
$\lambda$ satisfies (see \rref{eq:lambda})
\begin{eqnarray}\sum_{\ell=1}^{\nmin} \Lbr \log \lp \frac{1}{\lambda
|\lambda_\ell|^2} \rp \Rbr^+ = R(1+\epsilon).
\end{eqnarray}
Since the codeword differences satisfy the condition in
\rref{eq:mimo_univ_crit}, $\lambda$ can be explicitly calculated
(see \rref{eq:define_k} and \rref{eq:lambda_k}) \beq
\frac{1}{\lambda}  = 2^{R(1+\epsilon)}\left(|\lambda_1|\cdots
|\lambda_{\nmin}|\right)^{\frac{2}{\nmin}}. \eeq Thus the worst-case
pairwise error probability can be upper bounded by (see
\rref{eq:distance_k}): \beq\label{eq:emd} Q\lp\frac{\sqrt{\nmin
2^{R(1+\epsilon)}\left(|\lambda_1|\cdots
|\lambda_{\nmin}|\right)^{\frac{2}{\nmin}}- \sum_{\ell = 1}^{\nmin}
    |\lambda_\ell|^2}}{\sqrt{2}}\rp. \eeq Again using the
supposition in  \rref{eq:mimo_univ_crit}, the first term in
\rref{eq:emd} is growing unbounded with increasing SNR, while the
second term in \rref{eq:emd} is bounded above by $2n_tT$ (a
constant) because of the power constraint. Thus, the second term can
be ignored for increasing SNR and we can write the following upper
bound to the worst-case pairwise probability of error (using
\eqref{eq:mimo_univ_crit})
 \beq \nonumber Q\lp\frac{2^{R\epsilon}}{\sqrt{2}}\rp < \exp\lp
\frac{-2^{R\epsilon}}{2}\rp. \eeq With $R = r\log\SNR$, we conclude
that the pairwise error probability conditioned on the channel
realization not in $\mathcal{O}_\eps$ decays exponentially with SNR.
Since the number of codewords is  polynomial in $\SNR$, the overall
error probability conditioned on the channel realization not in
$\mathcal{O}_\eps$ decays exponentially with SNR. Thus the error
probability decays at the same rate as $\prob{\mathcal{O}_\eps}$.
Letting $\eps$ become arbitrarily close to zero, this decay rate can
be made arbitrarily close to that of the outage probability. Thus
the sequence of codes achieves the optimal tradeoff curve, and
further for every fading distribution, we conclude that the sequence
of codes satisfying \eqref{eq:mimo_univ_crit} is approximately
universal. This completes the sufficiency part of Theorem
\ref{thm:au_mimo}; necessity is proved in Appendix
\ref{ap:precise_char}. Next, we discuss some explicit schemes that
are approximately universal, starting with the simple scalar channel
and then moving onto more complex channel models.

\section{QAM is Approximately Universal for the Scalar Channel}\label{sec:scalar}
The single antenna (transmit and receive) channel model can be
written as (dropping the time index):
\begin{eqnarray*}
y & = & h x + w.
\end{eqnarray*}
The criterion for approximate universality (cf. Theorem
\ref{thm:au_mimo}) simply translates into a minimum distance one for
the code: \beq\label{eq:scalar-univ} \dmin^2  >
\frac{1}{2^{R(\SNR)+o\lp\log\SNR\rp}}, \eeq where $\dmin$ is the
normalized minimum distance over all the codeword pairs for the
coding scheme. Now, consider a simple coding scheme with unit block
length: QAM of size $2^R$. The normalized minimum distance of this
QAM has the property \beq\nonumber \dmin^2  \approx \frac{1}{2^R},
\eeq and is therefore approximately universal for the scalar fading
channel.

\section{The Parallel Channel}\label{sec:parallel-channel}
The parallel fading channel with $L$ diversity branches at time $m$
is \beq\label{eq:diversity_model_again} y_{\ell}[m] = h_{\ell}
x_{\ell}[m] + w_{\ell}[m], \qquad \ell =1, \ldots ,L. \eeq Here
$w_{1}[m],\ldots ,w_{L}[m]$  are i.i.d.\ $\CN\lp 0,1\rp$. The
approximate universality criterion for the parallel channel is
stated in the following theorem. The proof is very much similar to
the general approximate universality proof in
\ref{sec:proof_au_mimo}, hence we omit it here.
\begin{thm}\label{thm:sub}
A sequence of codes with rate $R(\SNR)$ bits/symbol is approximately
universal if and only if, for every pair of codewords, the
normalized codeword differences $\vd_1,\ldots ,\vd_L$ (the rows of
the difference codeword matrix) satisfy \beq\label{eq:par_univ_crit}
\|\vd_1\|^2\cdot\|\vd_2\|^2\cdots \|\vd_L\|^2  >
 \frac{1}{2^{R(\SNR)+o\lp\log\SNR\rp}}. \eeq
\end{thm}

In the rest of the section, we study a simple class of codes that
are approximately universal. Our main focus is on unit block length
codes based on permutations of a QAM  constellation that we call
{\em permutation} codes.\footnote{These codes are intimately related
to interleaver designs in turbo codes.} We show in
Section~\ref{sec:par} that even a random  permutation code is
approximately universal; thus space-only approximately universal
codes exist. Finally, we demonstrate simple examples of
approximately universal permutation codes: these codes are easy to
represent (so the storage complexity is low) and very easy to encode
and decode (so the run time complexity is small as well). The
parallel channel with two sub-channels is studied in
Section~\ref{sec:explicit} where a {\em bit-reversal permutation} is
shown to be approximately universal; this scheme also provides an
operational interpretation to the  outage condition (defined based
on an information theoretic underpinning) of the parallel  channel.
Simple permutation codes for the  parallel channel with more than
two sub-channels are the topic of
Section~\ref{sec:bit-generalization}.

\subsection{Approximate Universality of Codes Based on Rotation of PAM}

The criterion of maximizing the product-distance has been known in
the context of the i.i.d.\ Rayleigh fading channel. A code
construction based on rotations of PAM constellations is discussed
in \cite{BVRB96}: the transmit codeword vector $\bx :=
[x_1,\cdots,x_{L}]$ is defined as \beq\label{eq:bv98} \bx  =  \bu
\mM, \eeq where $u_1,\ldots,u_L$ are independent PAM constellations
and $\mM$ is an orthonormal matrix. \cite{BVRB96} shows existence
of $\mM$ such that the code has the maximum diversity possible, \ie
a non-zero product distance. The problem of explicitly maximizing
the minimum product distance was later considered in \cite{BV98}: it
was treated as an optimization problem over $\mM$ for fixed input
constellations. For $L=2$, the $\mM$ that maximizes the product distance
was explicitly found using computer simulations. Later,  a similar
idea of rotating QAM constellations was proposed in \cite{YW03} as a
part of the $2 \times 2$ code construction. It follows from Theorem
2 in \cite{YW03} that these codes are also in fact approximately
universal for the parallel channel.

Unfortunately, no generalizations of the rotation based codes exist
when there are more than two sub-channels. Further, these codes are
hard to decode  for large constellation sizes. Therefore, we propose
another approach: QAM constellations are the basis of the code
design but we consider mappings that utilize the {\em algebraic}
structure of the constellation; these mappings are {\em nonlinear}
with respect to the Euclidean vector space in which the QAM
constellations are embedded -- this is in contrast to the rotation
operation which is a linear mapping.

\subsection{Permutation Codes}\seclbl{par}
We would like to construct simple space-only (\ie unit block length)
approximately universal codes. As a step towards simple encoding and
decoding,  suppose the QAM constellation to be the alphabet for
each sub-channel. We need to protect every codeword by coding it
across every sub-channel: for the code to have any chance of  being
approximately universal, it should allow reliable communication for
every channel realization not in outage and, in particular, over the
parallel channel where all but one sub-channel is zero. Two design
implications are suggested:
\begin{enumerate}
\item With a rate of $R$ bits/symbol, each of the QAM
constellations on the sub-channels has $2^{R}$ points.
\item With $2^{R}$-point QAM  as the alphabet for each sub-channel, the points in the constellation over each
sub-channel can be identified one-one with points in the
constellation of the other sub-channels. In other words, the QAM
constellation over one sub-channel is a {\em permutation} of the
points in the QAM constellation over any other sub-channel.
\end{enumerate}
Mathematically, the permutation code can be represented as
\begin{eqnarray}
\nno \Co = \lbr \sqrt{\frac{\SNR}{2^{R}}}\lp
q,f_2(q),...,f_{L}(q)\rp| q \in \zQ\rbr,
\end{eqnarray}
where \beq \zQ = \lbr\lp a+ib\rp : -\frac{2^{\frac{R}{2}}}{2} \leq
a,b \leq \frac{2^{\frac{R}{2}}}{2}\rbr\eeq is the integer-QAM with
$2^{R}$ points, and $f_2,...,f_{L}$ are {\em
permutations} of $\zQ$. %We call such codes permutation codes.

\subsubsection{Examples}
\begin{figure}[h]
\begin{center}
\scalebox{0.8}{\input{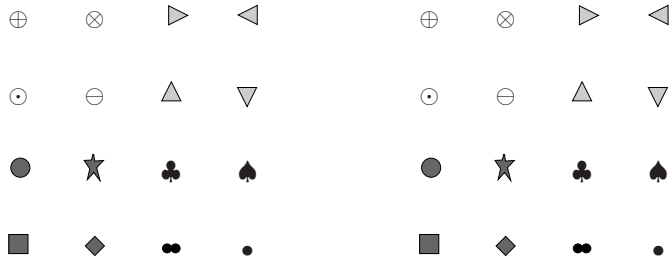}}
\end{center}
\caption{Repetition coding: $L = 2, ~R = 4$.} \label{fig:identity}
\end{figure}
Repetition coding is a simple example of a permutation code:  the
permutations are just the identity. \figref{identity} illustrates
the permutation code with identity permutation for $L=2$. Here $\zQ$
is the QAM with $16$ points.

\begin{figure}[h]
\begin{center}
\scalebox{0.8}{\includegraphics{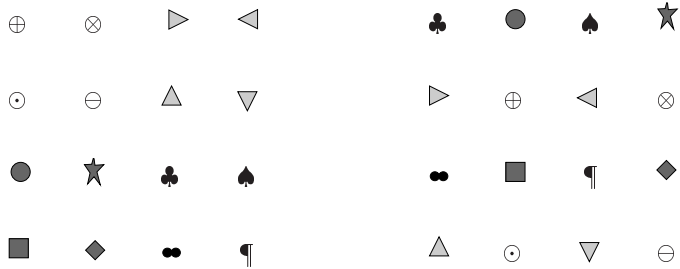}}
\caption{Permutation code: $L = 2,~ R = 4$} \label{fig:16qam}
\end{center}
\end{figure}
For $L =2$, \figref{16qam} shows a permutation code with 16
codewords that is designed to maximize the minimum product distance.
Product distance of this code is an improvement over the repetition
code in \figref{identity} by a factor of $4$. The code in
\figref{16qam} and its generalization to larger $L$ is discussed in
\cite{YPM04} using the theory of spreading transforms. The focus in
\cite{YPM04} is on finding codes that have a non-zero product
distance and can be efficiently constructed from smaller
constellations (QPSK) using spreading transforms.
%Though the
%techniques can be used for general QAM constellations,
%because of the QPSK based constructions, the rate of codes
%considered in \cite{YPM04} grows with $L$ but is fixed for a
%particular value of $L$. Here we consider high rate codes for a
%fixed value of $L$.

\subsubsection{A Random Permutation Code Ensemble}
Our search for permutation codes that are approximately universal
leads us  to study permutations with large QAM alphabet sizes. To
get a feel for whether there indeed exist permutation codes with
large enough product distance, we can look at an appropriate random
permutation ensemble and see if the product distance {\em averaged}
over this ensemble of permutation codes has the desired property. If
this is the case, then there must have been at least one permutation
code in the ensemble that is approximately universal. Averaging the
product-distance itself is not good enough; we look at the inverse
of the product distance and average it over all possible permutation
codes with the uniform measure. Our main result is the demonstration
of existence of permutation codes that are approximately universal:
\begin{thm}\label{thm:random_permutations}
There exists a sequence of permutation codes that is  approximately
universal over the parallel channel.
\end{thm}
The details of the proof are relegated to Appendix
\ref{ap:random_permutations}.

\subsection{Two Sub-channels: Bit-Reversal Permutation
Code}\label{sec:explicit} While it is encouraging to know the
existence of permutation codes that are approximately universal, it
is of engineering interest to actually construct simple
approximately universal codes from this ensemble. It turns out that
an {\em operational interpretation} of the outage condition (which
was defined based on an information theoretic understanding of the
compound channel) suggests natural permutation codes that are
approximately universal. In this section, we focus on the special
case when the parallel channel has just two sub-channels, \ie $L=2$.

\subsubsection{Operational Interpretation to the Outage Condition}
If we communicate at a rate of $R$ bits/symbol over the parallel
channel, the no-outage condition is \beq \label{eq:no-outage} \log (
1 + |h_1|^2\SNR) + \log (1+|h_2|^2\SNR)
> R. \eeq One way of interpreting this condition is as though the
first sub-channel provides $\log(1+|h_1|^2\SNR)$ bits of information
and the second sub-channel provides $\log(1+|h_2|^2\SNR)$ bits of
information, and as long as the total number of bits provided exceed
the target rate, then reliable communication is possible. In the
high SNR regime, we exhibit below a permutation code that makes the
outage condition concrete.

Suppose we independently code over the I and Q channels of the two
sub-channels. So we can focus on only one of them, say, the I
channel. We wish to communicate $R/2$ bits over two uses of the
I-channel. Analogous to the typical event analysis for the scalar
channel, we can exactly recover all the $R/2$ information bits from
the first I sub-channel  alone if: \begin{equation}\nonumber
\frac{1}{2}\log\left( 1 + |h_1|^2 \SNR \right)
> \frac{R}{2}. \end{equation}

However, we do not need to use just the first I sub-channel to
recover all the information bits:  the second I sub-channel  also
contains the same information and can be used in the recovery
process. Indeed, if we create $x_1^{\rm I}$ by treating the ordered
$R/2$ bits as the binary representation of the points $x_1^{\rm I}$,
then one  would intuitively expect that if
 \begin{equation}\label{eq:outage_k1} \frac{1}{2}\log\left( 1 + |h_1|^2 \SNR \right) > k_1, \end{equation} then one
should be able to recover at least $k_1$ of the most significant
bits of information. Now, if we create $x_2^{\rm I}$ by treating the
{\em reversal} of the $R/2$ bits as its binary representation, then
one should be able to  recover at least $k_2$ of the most
significant bits, if \begin{equation}\label{eq:outage_k2}
\frac{1}{2}\log\left( 1 + |h_2|^2 \SNR \right) > k_2.
\end{equation} But due to the reversal, the most significant bits in
the representation in the second I sub-channel are the least
significant bits in the representation in the first I sub-channel.
Hence, as long as $k_1+k_2 \ge R/2$, then we can recover {\em all}
$R/2$ bits. This translates to the condition
\begin{equation} \log (1+|h_1|^2\SNR) + \log (1+ |h_2|^2\SNR) > R, \end{equation}
which is precisely the no-outage condition (\ref{eq:no-outage}).
Thus, the bit-reversal scheme gives an operational meaning to the
outage condition.

\subsubsection{Bit-Reversal Permutation Code}
\begin{figure}[h]
\begin{center}
\input{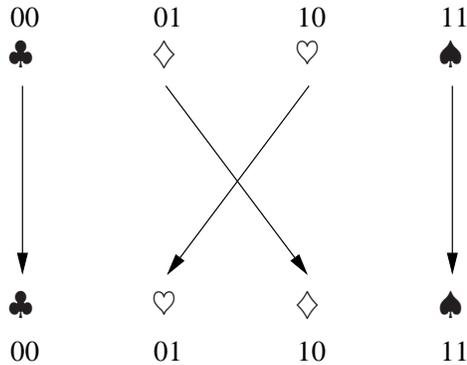}
\end{center}
\caption{The bit-reversal map for a 4-PAM.}\label{fig:bit-r}
\end{figure}

To make this idea concrete, first we need to define bit reversal. A
QAM can be thought of as two independent PAMs, and using I and Q
channels separately is equivalent to taking the QAM permutation as
two independent PAM permutations. Therefore we concentrate on one of
the PAMs and define the bit-reversal permutation for it. For a PAM
with $2^{R/2}$ points, we number the points from left to right by
$0$ to $2^{R/2}-1$. Based on this numbering, a canonical bit
sequence of length $R$ represents  each point in the PAM
constellation. Bit reversals are defined based on this
representation. The bit-reversal map for the  $4$-PAM is illustrated
in Figure~\ref{fig:bit-r}.

\subsubsection{Product Distance and Bit
Reversals}\label{sec:prodbit}
To show that the bit reversal scheme is approximately universal, we
have to show that it satisfies the criterion in
\rref{eq:par_univ_crit}. However, the plain bit-reversal is {\em
not} approximately universal. The problem is the inherent assumption
in the operational interpretation that if two points have different
MSB, then they are far apart geometrically and hence cannot be
confused with each other. This, however, is not true. Consider the
points with the binary representations:
\begin{eqnarray*} 011\cdots10 & \mbox{and} & 100\cdots01.
\end{eqnarray*}
Even though their MSB is different, they are separated by a fixed
distance of $3$ independent of the length $R/2$ of the binary
representation. The same is true for their bit-reversals. Thus, the
product distance between this codeword pair is $\frac{9}{2^{2R}}$
and it does not satisfy \rref{eq:par_univ_crit} for large $R$.

Even though the simple bit-reversal is not optimal, it can be
modified so that it essentially retains the operational
interpretation (so it is still easy to decode) and is approximately
universal. We discuss two such modifications here: irregularly
spaced PAM and alternate-bit-flipping.

\subsubsection{Irregularly Spaced PAM Permutation Code}
We have seen that the problem with the bit-reversal scheme is the
inherent assumption that the two points having different MSB are
geometrically far apart. A simple way to get around this problem is
to put gaps in the PAM constellation. That is, we introduce  a gap
of $g 2^{R/2}$ between $011\cdots1$ and $100\cdots0$ so that any two
points with different MSB are indeed far apart. More precisely, to
retain the operational interpretation, one has to put a gap of $g
2^{m}$ for every $m\th$ bit-change to ensure that the product
distance condition is met. The PAM constellation is now  {\em
irregularly spaced}.\footnote{The same idea of introducing  gaps  is
also present in the  Cantor set based representation in \cite{S01}.}

Consider any  two points in the irregularly spaced PAM
constellation. Suppose the first MSB they differ in their bit
representation is the $m\th$ one:  then by  construction the
normalized distance between the two points is lower bounded by
\beq\nonumber
 g  2^{-m}
\eeq The bit-reversals of these two points must have the same $m-1$
LSBs but a different  $m\th$ LSB; so the normalized distance between
the bit-reversals of these two points is lower bounded by
\beq\nonumber 2^{m-R/2}. \eeq Putting these two together, we
conclude that the normalized  product distance between a pair of
codewords in the bit-reversed irregularly spaced  permutation code
is lower bounded as
\begin{eqnarray*}
|d_1d_2| & \geq & g 2^{-m} 2^{m-R/2}  \\
& = & \frac{g}{2^{R/2}}.
\end{eqnarray*}
Comparing this with \rref{eq:par_univ_crit}, we conclude that the
code is  approximately universal.

A potential drawback of this approach is that the extra gaps
translate into an increase in the amount of power used for the same
rate. Thus, for a PAM of size $2^{R/2}$, the normalized increase in
size is given by
\begin{eqnarray*}
\sum_{m=1}^{R/2} g 2^{m-{R/2}} (\mbox{number of $m\th$ bit-changes}) &=&\sum_{m=1}^{R/2} g 2^{m-R/2} (2^{R/2-m}), \\
&=& g R/2 .
\end{eqnarray*}
With $R = r\log\SNR$, the SNR of this scheme is increased by a
factor of $\lp 1 + gr\log\SNR/2\rp$. In the diversity-multiplexing
scaling of our interest, this is an insignificant increase and thus
the code is still approximately universal.

\subsubsection{Alternate-Bit-Flipping Permutation Code}
\label{sec:altenatebitflip}
\begin{figure}[h]
\begin{center}
\input{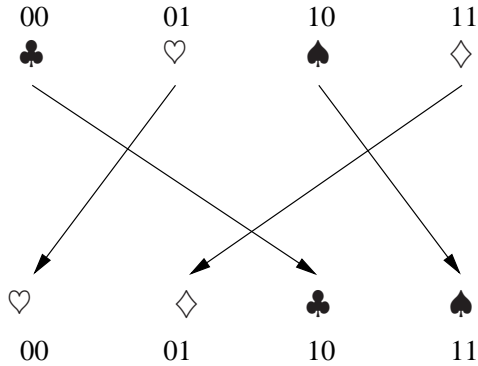}
\end{center}
\caption{Bit Reversals with alternate bits
  flipped.}\label{fig:alt-bit-r}
\end{figure}

Another modification of the plain bit-reversal scheme is to flip
every alternate bit after reversing. For example,  the
 point in the PAM constellation  with bit representation
%\begin{equation*}
111111
%\end{equation*}
is mapped to the point in the PAM constellation with bit
representation
%\begin{equation*}
010101.
%\end{equation*}
The scheme is illustrated for the 4-PAM constellation in
Figure~\ref{fig:alt-bit-r}.

\noindent In general, consider the $R/2$-bit representation of
integers $a_1$ and $a_2$ between $0$ and $2^{R/2}-1$:
\begin{eqnarray}\nno
a_1 & = & b_{R/2}^1 \cdots b_1^1,
\\\nonumber a_2 & = & b_{R/2}^2 \cdots b_1^2.
\end{eqnarray}
The the alternate-flip bit-reversal map $B$ is defined as (assuming
$R$ is even):
\begin{eqnarray}\nno
B(a_1) & = & \overline{b_1^1}b_2^1 \cdots
\overline{b_{R/2-1}^1}b_{R/2}^1,
\\\nonumber B(a_2) & = & \overline{b_1^2}b_2^2 \cdots \overline{b_{R/2-1}^2}b_{R/2}^2.
\end{eqnarray}
An easy observation is that this scheme  maintains the integrity of
the operational interpretation since the decoder can always flip the
bits back after estimating the flipped bits. Further, this scheme
turns out to be approximately universal:
\begin{thm}\label{thm:abf}
For every $a_1$ and $a_2$ between $0$ and $2^{R/2}-1$:
\bqa\label{eq:abfpd}
\frac{|a_1-a_2|}{2^{R/2}}\frac{|B(a_1)-B(a_2)|}{2^{R/2}} & \geq &
\frac{1}{8\cdot 2^{R/2}}. \eqa
\end{thm}
\noindent The details of the proof are somewhat involved and are
relegated to Appendix \ref{ap:abf}.

\subsection{Explicit Permutation Codes for General Parallel
  Channel}\label{sec:bit-generalization} In an effort to generalize
the bit-reversal scheme consider the following alternative, but
equivalent, view of the same scheme (for $L=2$).
\subsubsection{Bit-Reversal as a Linear Operation}
Each codeword in the bit-reversal permutation code is represented by
a sequence of, say  $2n$ bits. The first $n$ bits correspond to a
point in a  $2^n$-PAM constellation. The corresponding PAM
constellation point is then transmitted over the I channel of the
first sub-channel. The last $n$ bits similarly correspond to a point
in another $2^n$-PAM constellation which is then transmitted over
the Q channel of the first sub-channel. The transmissions over the I
and Q channels of the second sub-channel are the points in the PAM
constellation that correspond to bit-reversals of the first and last
$n$ bits, respectively, of the total $2n$ bits that define the
codeword.

If we fix the mapping between the sequence of bits and points in a
PAM constellation, the  bit-reversal scheme can be viewed entirely
as an operation on the $2n$ bits that represent the codeword.
Further more, if we  decide to do  the {\em same operation} over
both the I and Q channels (as in  the bit-reversal scheme), then we
only need  to consider operations over the first $n$ bits that
represent the codeword. In the rest of this discussion, we consider
only the operation on the first $n$ bits representing the codeword.
The operation involved in bit-reversal is particularly simple: it is
a  {\em linear} operation  on the vector of  bits (over
 the field $\mathbb{F}_2$). Linear operations can be represented
by matrices and the  bit reversal scheme corresponds to two
matrices: the {\em identity}  matrix ($\mI_{n}$) for the first
sub-channel and the {\em cross-diagonal} matrix with unit entries on
the cross diagonal ($\mD_n$) for the second sub-channel.

The outage interpretation implies that the decoder can deduce $k_1$
most significant bits from the first sub-channel (see
\rref{eq:outage_k1}) and $k_2$ most significant bits from the second
sub-channel (see \rref{eq:outage_k2}). Because of the simple
mappings in this case, the $k_1$ bits from the first sub-channel
correspond to the first $k_1$ bits of the vector of $n$ bits
representing the codeword and $k_2$ bits from the second sub-channel
that correspond to the last $k_2$ bits of the vector of $n$ bits
representing the codeword. As long as $k_1+k_2\geq n$, the decoder
can determine the codeword correctly.

\subsubsection{Universally Decodable Matrices}
This view of the bit-reversal scheme suggests a natural
generalization to  more than two sub-channels. We first generalize
the bit representation of the integers points of the PAM
constellation: we allow $q$-digit representation over a finite field
$\mathbb{F}_q$. Next we consider a (sequence of) collection of $L$
matrices $\lbr \mA_1^{\lp
  n\rp},\ldots ,\mA_L^{\lp n\rp}\rbr_n$ of size $n\times n$ with
entries selected from the finite field $\mathbb{F}_q$.  These
matrices naturally generate a sequence of permutation codes: for a
permutation code conveying $2n$ $q$-digits  of information, we
transmit over the I channel of the $\ell\th$ sub-channel the point
in the $2^n$-PAM constellation that corresponds to the $q$-digit
sequence that results from the linear operation of $\mA_{\ell}^{\lp
  n\rp}$ over the first $n$ $q$-digits of the $2n$ information
$q$-digits. This is done for each of the $\ell = 1,\ldots ,L$
sub-channels. Further, the same linear operations are used on the
last $n$ information $q$-digits to transmit points from the PAM
constellation on the Q channels of the $L$ sub-channels.

We say that this collection of matrices is {\em universally
decodable} if for any $k_1,\ldots ,k_L$ such that
\beq\label{eq:nooutage_implication} k_\ell \geq 0,\ell =1,\ldots ,L
\quad \mbox{and} \quad  \sum_{\ell
  =1}^L k_\ell \geq n,
\eeq the collection of the  first $k_1,\ldots ,k_L$ rows of the
matrices $\mA_1^{\lp n\rp},\ldots ,\mA_L^{\lp n\rp}$ respectively is
{\em full rank}, i.e., spans the vector space $\mathbb{F}_q^n$.

Universally decodable matrices (UDMs) provide an  operational
interpretation to the information theoretically defined  outage
condition. The number $k_\ell$ can be interpreted as the amount of
$q$-digits provided by the $\ell\th$ sub-channel; this depends on
the corresponding channel amplitude $|h_\ell|$. If the channel is
not in  outage, then \eqref{eq:nooutage_implication} holds. The full
rank condition implies that a unique codeword can be decoded
whenever the channel is not in outage. We formally state the
implication of this operational interpretation to outage below; the
proof is relegated to Appendix~\ref{ap:universal_decoding}.
\begin{thm}\label{thm:universal_decoding}
A sequence of UDMs leads to an approximately universal permutation
code sequence.
\end{thm}

Observe that the encoding and decoding complexity of the code based
on UDMs is simply {\em linear} in the number of bits $n$ and the
number of sub-channels $L$. The representation of the code involves
storing the $L$ matrices with a total of $Ln^2$ entries, again a
very small number.

In the rest of this subsection, we focus on explicit construction of
UDMs. First, we show how UDMs can be easily constructed from
maximum-distance separable codes (MDS) (though these constructions
require a field size that  grows  with $n$). In section
\ref{sec:udm3}, we present fixed field size constructions for $L=3$
and then discuss a  recent construction
\cite{VG05}, for arbitrary $L$.

\subsubsection{Reed-Solomon Codes are Approximately Universal} In
general, some
progress on the search for universally decodable matrices can be
made by strengthening the requirement on the collection of matrices
by requiring the collection of {\em any} $n$ rows from the matrix
\begin{eqnarray*}
\mA^{\lp n\rp} & = & \left[ \begin{array}{cccc} \mA_1^{\lp n\rp^t}
&\mA_2^{\lp n\rp^t} & \cdots & \mA_L^{\lp n\rp^t} \\ \end{array}
\right],
\end{eqnarray*}
to be full rank. Note that such a collection of matrices is still
universally decodable. This problem is same as designing a maximum
distance separable (MDS) codes with $\mA^{\lp n\rp}$ as its parity
check matrix. The condition universal decodability condition is the
same as requiring that the minimum distance of the code to be at
least $n+1$. Since $\mA^{\lp n\rp}$ is an $n \times Ln$ matrix, such
a code has length $Ln$ and rate $Ln-n$. A simple singleton bound
shows that then the code must be $[Ln, Ln-n, n+1]$\footnote{An
$[n,k,d]$ code over $\mathbb{F}_q$ is a linear, length-$n$ code with
$q^k$ codewords and a minimum Hamming distance of $d$. Its parity
check matrix is a $n-k \times n$ matrix over $\mathbb{F}_q$. Codes
for which $d=n-k+1$ meet the singleton bound (see Chapter 3.2 in
\cite{BLA03}) and are called MDS codes. These codes are well-studied
in coding theory and explicit codes like the Reed-Solomon codes are
MDS codes.}

Simple examples of such a code exist and this allows us to
explicitly construct the parity check matrix $\mA^{\lp n\rp}$. For a
finite field $\mathbb{F}_q$, a $[q+1,k,q-k+2]$ extended {\em
Reed-Solomon} code can be explicitly constructed for {\em every}
$k\leq q+1$ (see Chapter 6.8 of \cite{BLA03} for the exact parity
check matrix). For the extended Reed-Solomon codes, the field size
grows with the block-length. In fact, the field size is at least
$Ln-1$. In our setting, $n$ grows as $\log\SNR$, thus the field size
grows like $\log\SNR$. As noted in the proof of Theorem
\ref{thm:universal_decoding}, this still gives an approximately
universal code.

Next, we focus on the situation of practical and theoretical interest:
constructing UDMs with a  field size not
growing with $n$. With $L=2$, we
have already seen an example: $\lbr \mI_n,\mD_n\rbr_n$, where
$\mI_n$ is the $n\times n$ identity matrix and $\mD_n$ is the
$n\times n$ cross-diagonal matrix with all unit entries on the cross
diagonal; here the field size $q=2$.

\subsubsection{$L= 3$: Universally Decodable
Matrices}\label{sec:udm3}

Consider the following collection of binary matrices (\ie the field
size $q = 2$): $\lbr \mI_n,\mD_n,\mT_n\rbr$, where $\mI_n$ and
$\mD_n$ are, as before, the $n\times n$ identity and cross-diagonal
matrix with unit cross diagonal entries, respectively. $\mT_n$ is
defined using the recursive definition: \beq\label{eq:rec} \mT_{2n}
\df \left[\begin{array}{cc} \mT_n & \mT_n \\ {\bf 0} & \mT_n
\end{array} \right], \eeq with $\mT_1 = [1]$. Equivalently, $\mT_{2n} = \mT_2 \otimes
\mT_n$, where $\otimes$ denotes the {\em tensor} or {\em Kronecker}
product operation between two matrices (cf.\ Chapter~4.2 in
\cite{HJ91}). For $2^{m-1} < n < 2^m$, we define $\mT_n$ to be the
principal sub-matrix of $\mT_{2^m}$. We omit our original proof of
this result (it is still available in an earlier version of this
paper \cite{TV05-1}), in light of a crisper proof  that follows from
a more general result in  \cite{VG05}; this generalization was
motivated by the present construction for $L=3$.

For $L=4, q=3$ computer simulations are used in \cite{Dos05} to
justify the conjecture that the following collection of matrices is
universally decodable: $\lbr \mI_n,\mD_n,\mT_n,\mR_n\rbr_n$ where
the first two matrices are, as before, the $n\times n$ identity and
cross-diagonal matrix with unit cross diagonal entries,
respectively.  With $n=3$, define \beq
 \mT_3 \df \left[\begin{array}{ccc}
1 & 2  & 1\\
0& 1 & 1 \\
0 & 0 & 1
\end{array}\right] \quad \mbox{and}\quad
\mR_3 \df \left[\begin{array}{ccc}
1 & 1  & 1\\
0& 1 & 2 \\
0 & 0 & 1
\end{array}\right].
\eeq For $n$ a power of 3, we define, recursively, $\mT_{3n} = \mT_3
\otimes \mT_n$ and $\mR_{3n} = \mR_3 \otimes \mR_{n}$, with the
multiplication operations in the context of  the field ${\mathbb
F}_3$. For $3^{m-1} < n < 3^m$, we define $\mT_n$ and $\mR_n$ to be
the principal sub-matrices of $\mT_{3^m}$ and $\mR_{3^m}$,
respectively. This conjecture has now been verified as a special case
of the general result in \cite{VG05}.

\subsubsection{A Complete Characterization of UDMs}
Motivated by the results in the previous two subsections, the authors
in \cite{VG05}, have recently completely solved the problem of
constructing UDMs.  They show for any $n$ the
condition $L \leq q+1$ is both necessary and sufficient. They
construct UDMs based on {\em Pascal's triangle}. We state their
construction  (see Proposition 9, \cite{VG05}), for completeness:
\begin{thm}
Let $q$ be a prime power and let $L \leq q+1$. Suppose $\alpha$ is a
primitive element over ${\mathbb F}_q$. Then the following matrices are
UDMs:
\begin{eqnarray*}
\mA_1 & = & \mI_n,\\
\mA_2 & = & \mD_n,\\
\left[\mA_{\ell}\right]_{(j,k)} & = &
\left(^k_j\right)\alpha^{(\ell-2)(k-j)}, ~\mbox{for}~1 \leq j,k \leq
n~\mbox{and}~3\leq \ell \leq L,
\end{eqnarray*}
where $(^k_j)$ is defined as the natural mapping to prime subfield
of ${\mathbb F}_q$ of the natural number
\begin{eqnarray*}
\left(^k_j\right) & := & \frac{k(k-1)\cdots (k-j+1)}{j(j-1)\cdots 1}.
\end{eqnarray*}
\end{thm}

\section{The MISO Channel}\label{sec:miso}
The parallel channel allowed us to study approximately universal
codes on channels with solely multiplexing gain. We now turn to
study channels that offer solely diversity gain: the MISO and SIMO
channels, with multiple transmit (receive) and single receive
(transmit) antennas, respectively. The SIMO channel can be reduced
to a scalar channel by considering a scalar sufficient statistic:
receive beamformed vector. Therefore, any approximately universal
scheme for the scalar channel, such as the QAM scheme (see Section
\ref{sec:scalar}), will also be approximately universal for the SIMO
channel. In this section, we focus on the MISO channel and
understand properties of approximately universal codes over this
channel.

The scalar output of a MISO channel with $n_t$ transmit antennas at
time $m$ can be written as \beq\nonumber y[m] =\vh^{t}\bx[m] +
w[m],\eeq where $\bx[m]$ is an $n_t$ dimensional vector input and
$\vh$ is the $n_t$-dimensional vector of fading gains $h_i$s.

\subsection{Characterization of Approximately Universal Codes}\label{sec:univ_miso}
The approximate universality criterion for the MISO channel can be
stated as (see Theorem \ref{thm:au_mimo}), for every codeword
difference matrix: \beq\label{eq:universal_miso_crit} \lambda_1^2
> \frac{1}{2^{R(\SNR)+o\lp\log\SNR\rp}}, \eeq where $\lambda_1$ is
the minimum singular value of the codeword difference matrix.

There is an intuitive explanation for this result: a universal code
has to protect itself against the worst channel that is not in
outage. The condition of no-outage  only puts a constraint on the
{\em norm} of the channel vector $\vh$ but not on its direction. So,
the worst channel aligns itself to the ``weakest direction'' of the
codeword difference matrix. The corresponding worst-case pairwise
error probability is governed by the smallest singular value of the
codeword difference matrix.

On the other hand, the i.i.d.\ Rayleigh channel does not prefer any
specific direction: thus the design criterion tailored to its
statistics requires that the {\em average} direction be well
protected and this translates to the determinant criterion. While
the two criteria are different, codes with large determinant tend to
also have a  large value for the  smallest singular value; the two
criteria (based on worst-case and average-case) are related in this
aspect.

For the case when $n_t=2$, the Alamouti scheme \cite{A98} converts
the MISO channel to a scalar channel with gain $\|\vh\|$ and the
{\em total} SNR reduced by a factor of 2. Hence, the outage behavior
is exactly the same as in the original MISO channel, and the
Alamouti scheme provides a {\em universal} conversion of the
$2\times 1$ MISO channel to a scalar channel. Any approximately
universal scheme for the scalar channel, such as a QAM, when used in
conjunction with the Alamouti scheme will be approximately universal
for the MISO channel.

In the general case when the number of transmit antennas is greater
than $2$, there is no equivalent to the Alamouti scheme. Here we
explore one approach to construct approximately universal schemes
for the general MISO channel: we consider a simple scheme that
converts the MISO channel into a parallel channel and show that the
scheme is approximately universal over a restricted class of MISO
channel statistics.

\subsection{MISO channel viewed as a Parallel Channel}
Consider the  simple scheme of using one antenna at a time to
communicate at a rate of $R$ bits/symbol on the MISO channel. By
using one transmit antenna at a time, we arrive at a parallel
channel with $n_t$ sub-channels and the data rate of communication
is $R$ bits/symbol per sub-channel. We code over the antennas using
a parallel channel code, e.g.\ a permutation code. Our first result
is that this simple scheme is tradeoff optimal for the i.i.d.\
Rayleigh fading MISO channel.
%; we state this formally below and delegate the proof to
%Appendix~\ref{ap:miso}.
%\begin{prop}\label{prop:miso_1}
%Using an approximately universal parallel channel code sequence over
%each of the antennas, one at a time, of an i.i.d.\ Rayleigh fading
%MISO channel is tradeoff optimal.
%\end{prop}

Can this conversion be approximately universal?  To see that this
could not be the case, consider the following (worst-case) MISO
channel model: the channels from all but the first transmit antenna
are very poor. To make this example concrete, set $h_{\ell} = 0, \,
\ell =2,\ldots ,n_t$. The tradeoff curve depends on the  outage
probability  (which depends only on the statistics of the first
channel). Using one transmit antenna at a time is a waste of degrees
of freedom: since the channels from the all but the first antenna
are zero, there is no point in transmitting any signal on them. Thus
the scheme could not have been tradeoff optimal over a MISO channel
with such statistics.

Essentially, using one antenna at a time equates temporal degrees of
freedom  with  spatial ones. All temporal degrees of freedom are the
same, but the spatial ones need not be the same: in the extreme
example above, the spatial channels from all but the first transmit
antenna are zero. Thus, it seems reasonable that when all the
spatial channels are {\em symmetric} then the parallel channel
conversion of the MISO channel is tradeoff-optimal. This intuitive
argument is formalized in the proposition below; the proof is
provided in Appendix \ref{ap:miso}.
\begin{prop}\label{prop:miso_2}
An approximately universal parallel channel code sequence used over
the antennas of a MISO channel, one antenna at a time, is
tradeoff-optimal for the class of MISO channels with i.i.d.\ fading
coefficients. Further, the optimal tradeoff curve of the MISO
channel is given by \beq\label{eq:miso-tradeoff} d^*(r) = an_t(1-r),
\quad 0\leq r \leq 1, \eeq where \beq a \df \lim_{x \rightarrow 0}
\frac{\log\mathbb{P}\lp |h_\ell|^2 \leq x \rp}{\log x}, \quad
\forall \ell = 1,\ldots ,n_t. \eeq
\end{prop}

\begin{figure}
\begin{center}
\epsfig{file=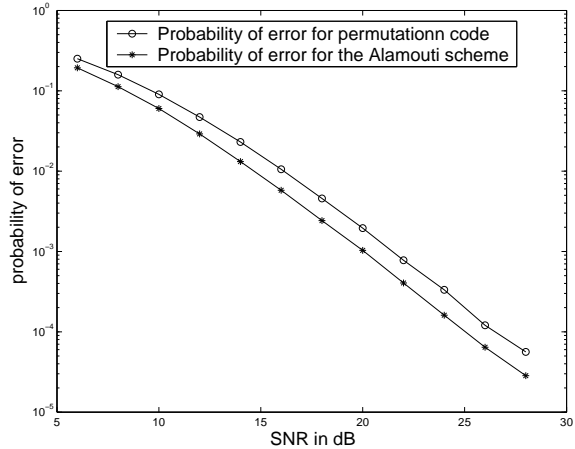, width=3in}
\end{center}
\caption{The error probability of  uncoded QAM with the Alamouti
scheme and that of a  permutation code over one antenna at a time
for the Rayleigh fading MISO channel with two transmit antennas: the
permutation code is only about 1.5 dB worse than the Alamouti scheme
over the plotted error   probability range.}
\label{fig:alamouti_permutation_compare}
\end{figure}

We have seen that the conversion of the MISO channel into a parallel
channel is tradeoff-optimal for the i.i.d.\ Rayleigh fading channel.
To get a practical feel for  how much loss the conversion of the
MISO channel into a parallel channel entails with respect to the
optimal outage performance, we plot the error probabilities of  two
schemes with the same rate ($R = 2$ bits/symbol): uncoded QAMs over
the Alamouti scheme and the permutation code in
Figure~\ref{fig:16qam}. This performance is plotted in
Figure~\ref{fig:alamouti_permutation_compare} where we see that the
conversion of the MISO channel into a parallel channel entails a
loss of about 1.5 dB in SNR for the same error probability
performance. This is a fairly small loss and suggests the practical
utility of the conversion of the MISO channel with larger number of
receive antennas to a parallel channel.

\section{The MIMO Channel}\label{sec:mimo}
Having studied the construction of approximately universal codes over
the parallel and
the MISO channel, we are now ready to move over the general
MIMO channel: we first  conclude the approximate
universality of some  recently proposed codes and then explore the
approximate universality properties of two classical space time coding
architectures: D-BLAST and  V-BLAST.

\subsection{Approximate Universality of Number-Theoretic Codes}
Some of the recent space time code constructions in the literature
have a number-theoretic flavor. In particular, a rotated QAM
constellation was used to construct a  two transmit antenna space
time code in \cite{YW03,DV03,BRV04}. For arbitrary $n_t$,
\cite{EKPKL04} proposes codes derived from cyclic division algebras.
Some constructions based on cyclic division algebras are also
presented in \cite{ORBV04, KR05}. All these two codes satisfy the
non-vanishing determinant criterion. The authors in \cite{YW03,DV03}
used this property to conclude the tradeoff optimality over the
i.i.d.\ Rayleigh fading channel. In the light of our
characterization of approximate universality (cf.\
Theorem~\ref{thm:au_mimo}), we can conclude that all these codes are
approximately universal; further more, in the light of the
discussion in Section~\ref{sec:dl_universality}, we can conclude
that these codes are approximately universal simultaneously for
every MIMO channel with $n_t$ transmit antennas  ($n_t=2$ for the
code in \cite{YW03,DV03}) and arbitrary $n_r$. To see this formally,
we discuss the two transmit antenna code in \cite{YW03} in some
detail.

The rotated code QAM code in \cite{YW03} spans two symbols and is
designed to work over the two transmit MIMO channel. The entries of
the $2\times 2$  transmit codeword matrix $\mX \df \Lbr x_{ij}\Rbr$
are \beq\label{eq:yao_wornell_code} \Lbr\begin{array}{l}x_{11}
\\x_{22}\end{array}\Rbr \df  \mQ(\theta_1) \Lbr\begin{array}{l}u_1 \\
u_2\end{array}\Rbr,\quad \mbox{and}\quad \Lbr\begin{array}{l}x_{21}
\\x_{12}\end{array}\Rbr \df  \mQ(\theta_2) \Lbr\begin{array}{l}u_3 \\
u_4\end{array}\Rbr. \eeq Here $u_1,u_2,u_3,u_4$ are independent QAMs
of size $2^{R/2}$ each (so the  data rate of this scheme is  $R$
bits/symbol). The rotation matrix $\mQ(\theta)$ is \beq\nonumber
\mQ(\theta) \df \left [ \begin{array}{cc} \cos \theta & -\sin \theta
\\ \sin \theta & \cos \theta \end{array} \right].
\eeq With the choice of the angles $\theta_1,\theta_2$  equal to
$1/2\tan^{-1}2$ and $1/2\tan^{-1}(1/2)$ radians respectively,
Theorem~2 of \cite{YW03} shows that the determinant of every
normalized codeword difference matrix $\mD$ satisfies \beq \nonumber
|\det\mD|^2 \geq \frac{1}{10\cdot 2^R}. \eeq Our discussion so far
is summarized in the following formal statement characterizing of
the performance of this code.
\begin{prop}
The code described in \eqref{eq:yao_wornell_code},  with $\theta_1 =
1/2\tan^{-1}2$ and $\theta_2= 1/2\tan^{-1}(1/2)$, is approximately
universal for every MIMO channel with two transmit antennas.
\end{prop}

\subsubsection{Discussion}
While  the two codes discussed above are explicit and easy to
encode, they lack a computationally simple decoding algorithm. In
general, it appears hard to design explicit approximately universal
codes for the MIMO channel with a computationally simple decoding
algorithm; it still remains an open problem. For the parallel
channel we have been able to answer this question to a reasonable
extent. The difference in the two models arises due to the {\em
rotation matrix} in the SVD decomposition \rref{eq:svd}: a parallel
channel code has to be optimal for a fixed rotation matrix (the
identity matrix) while a MIMO channel code has to be optimal for
{\em every} rotation matrix. This difference seems to naturally lead
to codes with a number-theoretic flavor: they are delicately
designed so as to cope with every possible rotation. Such a code
with a computationally simple decoding algorithm has not yet been
found.

An alternate view point is proposed in  \cite{GCD04} where a lattice
based space-time code  is constructed. The authors show that the
structure of these codes resembles random Gaussian codes and then
conclude the tradeoff optimality of an ensemble of lattice codes for
a decoder based on a generalized MMSE estimator for the i.i.d.\
Rayleigh fading channel. A typical code in this ensemble is very
unlikely to be approximately universal. In fact, one of the
important conclusions of the  the authors of \cite{GCD04} is that
their construction shows that maximizing the determinant criterion
is not a necessary requirement for achieving the tradeoff for {\em
specific} fading distributions. However, as we see here, maximizing
the determinant criterion is a {\em necessary and sufficient}
condition to design robust codes that are tradeoff-optimal for {\em
every} fading distribution.

\subsection{The V-BLAST Architecture}\label{sec:vb_universality}
The V-BLAST architecture was proposed for high rate communication
over the MIMO channel \cite{FGRW99}. It splits the data stream into
independent streams that are sent over the different transmit
antennas. It is very clear that V-BLAST is not tradeoff optimal at
low rates: the largest diversity of any data stream is limited by
the number of receive antennas. However, it is also clear that the
V-BLAST scheme cannot be approximately universal even at  high
rates: over the $2\times 1$ MIMO channel suppose  the channel from
one of the transmit antennas is zero and the other channel is
$\CN(0,1)$. Then the diversity obtained by the data stream sent over
the first transmit antenna for any multiplexing gain is zero whereas
the overall channel has a non-zero diversity-multiplexing tradeoff.
Since the V-BLAST scheme does not code across the transmit antennas
it takes a hit when the transmit antennas have asymmetric fading
statistics. When all transmit antennas are statistically similar to
one another, V-BLAST indeed turns out to be tradeoff optimal at high
rates; we explore this aspect in detail in Section~\ref{sec:vblast}.

\subsection{The D-BLAST Architecture}\label{sec:db_universality}
The D-BLAST architecture has been proposed to attain high diversity
gains over the MIMO channel \cite{F96}. The data is split into
independent streams that are sent over the MIMO channel in a
diagonal fashion. The coding scheme can be written as
\bq\label{eq:pdblast} \Lbr
\begin{array}{ccccccc}
0 & \cdots & 0 & p^{(1)}_{1} & p^{(2)}_{1} & \cdots & p^{(T-n_t+1)}_1 \\
\vdots & \adots & \adots & \adots & \cdots & \adots & \vdots\\
0 & p^{(1)}_{n_t-1}& p^{(2)}_{n_t-1} & \adots &  & \adots & 0 \\
p^{(1)}_{n_t}& p^{(2)}_{n_t} & \cdots & \cdots &  & 0 & 0
\end{array} \Rbr,
\eq where $\bp^{(k)}=\Lbr p_1^{(k)},\ldots,p_{n_t}^{(k)}\Rbr$ are
the independent data streams.

It is well known that the D-BLAST architecture with MMSE-SIC
receiver preserves mutual information over any deterministic MIMO
channel with Gaussian inputs; thus it converts a  MIMO channel into
an equivalent parallel channel (a tutorial description of this
conversion is described in Chapter~8.5 of \cite{TV05}). Therefore an
approximately universal code over the parallel channel, such as the
permutation code, when used as the streams of the D-BLAST
architecture for the MIMO channel will be approximately universal
for the MIMO channel. This approach of converting the MIMO channel
into a parallel channel has also been used by Matache and Wesel in
\cite{MW03}.

Alternatively, one can see its approximate universality  by
explicitly verifying that it satisfies the condition in
\eqref{eq:mimo_univ_crit} for $n_t = n_r$. The product of singular
values of the codeword difference matrix for \rref{eq:pdblast} turns
out to be lower bounded by the product distance of the permutation
code. Thus, if $\bp^{(k)}$ is a permutation code that is
approximately universal for the parallel channel, then the D-BLAST
scheme \rref{eq:pdblast} is approximately universal for the MIMO
channel (see and compare \rref{eq:par_univ_crit} and
\rref{eq:mimo_univ_crit}).

A potential  drawback is the initialization loss due to the zero
padding in \rref{eq:pdblast} which reduces the effective rate. For a
$2 \times 2$ channel with block-length three, a rate of $R$
bits/stream corresponds to a rate of $2R/3$ bits/symbol on the MIMO
channel. In general, the actual tradeoff curve achieved by this
scheme is
\begin{equation}\label{eq:finite_performance_dblast}
d_{\rm out}\lp \frac{T}{T-n_t+1}r\rp,
\end{equation}
where $r$ is the  multiplexing gain per symbol. For the block length
$T$ large, D-BLAST approaches approximate universality. For finite
block-length, this scheme is strictly sub-optimal. The precise
characterization for approximate universality also implies that this
performance can not be universally improved upon using a better
decoding strategy (than MMSE and successive interference
cancelation). In Section \ref{sec:dblast}, we see that the
performance can indeed be improved upon for a certain {\rm
restricted} class of fading distributions using a better decoding
strategy.

\section{The  V-BLAST Architecture}\label{sec:vblast}
The V-BLAST architecture transmits independent data streams over the
transmit antennas. This is closely related to how a multiple access
channel is operated, the tradeoff performance of which under i.i.d.\
Rayleigh fading  is studied (using random Gaussian codes) in
\cite{TVZ04,PV04}. In this section, we study the performance of
simple modulation schemes over the V-BLAST architecture: in
particular, QAM constellations. While we have seen that the V-BLAST
architecture can never  be approximately universal, it still
performs very well for an interesting {\em restricted} class of
channels.

\subsection{Tradeoff Optimality over Rayleigh Fading Channels}
\begin{figure}[htb]
\begin{center}
\scalebox{0.8}{\input{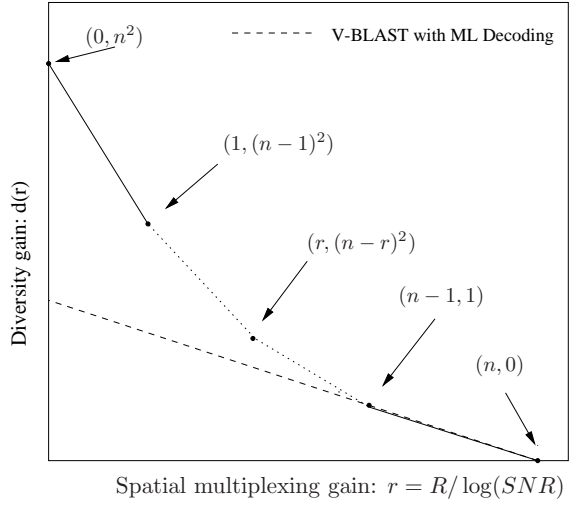}}
\end{center}
\caption{The i.i.d.\ Rayleigh fading channel with
$n_t=n_r=n$.}\label{fig:nn_tradeoff}
\end{figure}

Consider operating the V-BLAST architecture over  an $n_t\times n_r$
i.i.d.\ Rayleigh fading channel: we transmit independent data
streams over each of the $n_t$ antennas; each data stream is
transmitted {\em un-coded} using a  QAM constellation (with
$\SNR^{r/n_t}$ points at each time symbol). This scheme corresponds
to a total data rate of $r\log\SNR$ bits/symbol over the MIMO
channel. Our main result is the precise characterization of the
tradeoff performance; the proof is available in
Appendix~\ref{ap:qam}.
\begin{prop}\label{prop:vblast_rayleigh}
Uncoded independent QAMs of size $\SNR^{r/n_t}$ points over the
antennas of an $n_t\times n_r$ i.i.d.\ Rayleigh fading MIMO channel
are protected by a diversity gain, $d(r)$, where
\bqa\label{eq:performance_vblast} d(r) & = &  n_r  - \frac{n_r
r}{n_t} \quad {\rm if}~n_r \geq n_t
\\ & \geq & n_r - r \quad {\rm if}~ n_r < n_t.
\eqa
\end{prop}
Several interesting observations follow from this result.
\begin{enumerate}
\item Apart from the fact that the channel can be in outage, there is
  an additional error event in the V-BLAST architecture: the
  presence of the other simultaneously transmitted streams impacts
  the reliable reception of any particular data stream. However, the
  reliability performance  represented in
  \eqref{eq:performance_vblast} is as if the other streams didn't
  exist at all. This suggests that the typical way error occurs is
  not due to the inter-stream interference but because of the
  channel being in outage.
\item With $n_t = n_r = n$, the diversity gain of uncoded QAMs is  equal to $n - r$;
  this matches the optimal diversity gain  characterized in
  \cite{ZT03} for large enough $r$ ($\geq n-1$). This observation is
  graphically illustrated in Figure~\ref{fig:nn_tradeoff}.
\item In a multiple access setting with
\begin{itemize}
\item $n_t$ users with one transmit antenna each,
\item a symmetric multiplexing gain of $r/n_t$ per user,
\item $n_r \geq n_t$ receive antennas,
\end{itemize}
the diversity-multiplexing tradeoff is given by \cite{TVZ04}:\beq
n_r - \frac{n_r r}{n_t}. \eeq Therefore this simple scheme is
tradeoff-optimal.

\item With $n_t \neq n_r$, the performance of uncoded QAMs is never
  equal to the optimal diversity gain of the channel.
\end{enumerate}
Rayleigh fading is a physically relevant fading model and we have seen
the tradeoff optimality at high rates of plain uncoded QAMs using the
V-BLAST architecture. We can conclude the robustness of this
performance if it continues to hold for a wider class of fading
distributions; this is the focus of the next section.

\subsection{Tradeoff Optimality over Isotropic Fading Channels}

The key property of a fading distribution determining the diversity
performance is the {\em near zero} behavior of its singular values.
In particular, denoting $\phi_1,\ldots ,\phi_{\nmin}$ to be the
increasingly ordered {\em squared} singular values  of $\mH$,
suppose \beq\label{eq:decayrate_singularvalues} \prob{\phi_1 \leq
\epsilon_1,\ldots ,\phi_{\nmin}\leq \epsilon_{\nmin}}
\stackrel{.}{=}
 \epsilon_1^{k_1+1}\cdots\epsilon_{\nmin}^{k_{\nmin}+1}, \eeq for
$\eps_1 < \cdots < \eps_{\nmin}$. Here our notation
$f(\epsilon_1,\ldots,\epsilon_{\nmin}) \stackrel{.}{=}
g(\epsilon_1,\ldots ,\epsilon_{\nmin})$ is in the sense of \beq
\lim_{\eps_{1}\rightarrow 0} \lim_{\eps_{2}\rightarrow 0}\cdots
\lim_{\epsilon_{\nmin}\rightarrow 0} \frac{\log
f(\epsilon_1,\ldots,\epsilon_{\nmin})}{\log g(\epsilon_1,\ldots
,\epsilon_{\nmin})} = 1. \eeq We also assume that all the singular
values have an exponential tail, \ie, for there exists an $\eps$
such that for large enough $x$, \beq \prob{\phi_\ell \geq x} \leq
e^{-\eps x} \quad\forall\quad\ell.\eeq For a given near zero
behavior of singular values, the tradeoff curve can be explicitly
determined. We compute it for the case when $k_i$s are increasingly
ordered (as is the case for i.i.d.\ Rayleigh fading).

\begin{thm}\label{thm:general_tradeoff}
If $k_1 < k_2 < \cdots < k_{\nmin}$, then the tradeoff curve is
piecewise linear with $\nmin$ segments and the $s\th$ segment (\ie
$s \leq r < s+1$) is given by: \bqa\nno (k_{\nmin-s}+1)(s+ 1 -r) +
\sum_{\ell=\nmin-s+1}^{\nmin} \lp k_\ell + 1  \rp\eqa Furthermore,
random Gaussian codes with block-length $T \geq k_{\nmin-s}+1$ will
achieve this performance.
\end{thm}
\begin{proof} See Appendix \ref{ap:general_outage} for the outage curve calculation.
The proof of achievability for random Gaussian codes is a simple
generalization of achievability proof in \cite{ZT03} and we omit it
here.
\end{proof}

%For the i.i.d. Rayleigh fading, \ie $k_i = |n_r - n_t| + 2(i-1)$,
%this curve matches the piecewise linear curve joining points
%$(s,(n_r-s)(n_t-s)$ for $s = 0,\ldots,\nmin$.

%We formally state the performance in the  theorem below, the proof
%of which is delegated to Appendix~\ref{ap:general_outage}.
%\begin{thm}
%Consider transmitting uncoded QAMs with $\SNR^{r/n}$ points over an
%isotropic $n\times n$ MIMO channel $\mH$  with the polynomial decay
%rates of the ordered squared singular values defined in
%\eqref{eq:dcayrate_singularvalues}
%and denoted by $k_1,\ldots ,k_n$.
%Then, the diversity gain seen by the uncoded QAMs is equal to
%\begin{quote}
%{\tt formally write the gain in terms of $k_1,\ldots ,k_n$. Can
%  you
%write this result  for general $n_t,n_r$? If so, please do that by
%changing the premise of the theorem slightly. }
%\end{quote}
%\end{thm}
%This is the natural generalization of
%Proposition~\ref{prop:vblast_rayleigh}.

The key property of the i.i.d.\ Rayleigh fading channel used in the
calculation of the performance of uncoded V-BLAST transmission is
the rotational symmetry of its  statistics. We can thus generalize
this calculation and characterize the performance of uncoded V-BLAST
transmission over {\em isotropic} distributions   on the $n \times
n$ MIMO channel $\mH$: \beq\label{eq:isotropic_definition}
\mbox{$\mH\mQ$ has the same distribution as $\mH$ for every unitary
matrix $\mQ$}. \eeq

If the ordered singular values of the $n\times n$ MIMO channel $\mH$
decay {\em slower} than the corresponding decay rate of ordered
singular values of $\mH$ with   i.i.d.\ Rayleigh fading, then we can
extend our earlier observation of tradeoff optimality of the
transmission of uncoded QAMs over the  V-BLAST architecture at
multiplexing gains $r \geq n-1$ on the i.i.d.\ Rayleigh fading
channel. We make this precise in the following proposition,
delegating the proof to Appendix~\ref{ap:mimo_isotropic}.

\begin{prop}\label{eq:optimality_vblast_isotropic}
Consider $n\times n$ isotropic MIMO channels with the polynomial
decay rates of its squared singular values as defined in
\eqref{eq:decayrate_singularvalues}.  The uncoded QAM transmission
over the V-BLAST architecture at multiplexing rates $r \geq n-1$ is
tradeoff optimal for every isotropic MIMO channel satisfying
\begin{eqnarray*}
k_i & > & (2i-2)+k_1 \quad i=2,\ldots,n, \\\nno k_1 & \leq & 0.
\end{eqnarray*}
\end{prop}

\section{The D-BLAST Architecture}\label{sec:dblast}
We have seen (cf.\  Section~\ref{sec:db_universality}) that
the D-BLAST architecture with approximately universal parallel channel
codes over its independent constituent data streams approaches
approximately universality for large block length (cf.\
\eqref{eq:finite_performance_dblast}).  For any finite block length,
the architecture is strictly  {\em not}
approximately universal.  However, we will see in this section that by
restricting the class of MIMO channels over which we demand
universality, the performance of the D-BLAST architecture can be
significantly improved. In particular, our focus throughout this
section is with isotropic MIMO channels. We characterize the diversity
performance of the D-BLAST architecture with exactly two data streams;
our main result is the observation of a restricted
universality result for channels with $2$ receive antennas.

The i.i.d.\ Rayleigh fading MIMO channel is also isotropic and we state
our results first in this context; the calculations are relatively
simple and shed insight as to why we can expect robustness when
generalized to arbitrary isotropic channel distributions.

\subsection{Tradeoff Optimality over Rayleigh Fading Channels}
\begin{figure}
\begin{center}
\scalebox{0.8}{\input{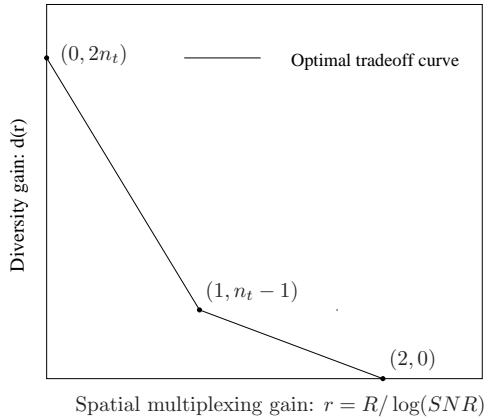}} \caption{The tradeoff
behavior for the $n_t\times 2$ i.i.d.\ Rayleigh
  fading channel.}
\label{fig:nr=2tradeoff}
\end{center}
\end{figure}

Consider the $n_t\times 2$ i.i.d.\ Rayleigh fading MIMO channel: the
tradeoff curve is composed of two linear segments, as illustrated in
Figure~\ref{fig:nr=2tradeoff}.

\subsubsection{D-BLAST and the First Segment}
Consider  the D-BLAST architecture with only two
independent data streams:
\begin{eqnarray}\label{eq:dblastnt2}
\Lbr \begin{array}{ccccc}
0 & \cdots & 0 & p_{n_t} & q_{n_t}  \\
\vdots & \adots & \adots & \adots & 0 \\
0 & p_2 & \adots & \adots & \vdots \\
p_1 & q_1 & 0 &\cdots & 0
\end{array} \Rbr;
\end{eqnarray}
here $\left[p_1, \ldots, p_{n_t} \right]$ and $\left[q_1,
\ldots, q_{n_t} \right]$ are unit block-length approximately universal
codes for a parallel channel with $n_t$ sub-channels. Suppose both
these codes have a data rate of
\beq\label{eq:singlestreamrate}
\frac{\lp n_t+1\rp r}{2n_t}\log\SNR \quad \mbox{bits per
  sub-channel}.
\eeq
Since the overall architecture is composed of two data streams and the
transmission lasts $n_t+1$ time symbols long, the overall data rate of
the  architecture is $r\log\SNR$ bits/symbol.
Our main result is a precise characterization of the diversity
performance  under joint ML decoding of the streams; the proof is
available in Appendix~\ref{ap:dblast}.
\begin{prop}\label{prop:dblast_firstsegment_rayleigh}
The D-BLAST architecture in \eqref{eq:dblastnt2} with approximately
universal parallel channel codes as its two data streams operated at
a total multiplexing gain of $r$ over the i.i.d.\ Rayleigh fading
$n_t\times n_r$ MIMO channel with $n_r \geq 2$ sees a diversity gain
equal  to \beq\label{eq:nt2_firstsegment}
 n_r\lp n_t - \frac{n_t + 1}{2} r\rp.
\eeq
\end{prop}

A couple of observations follow:
\begin{enumerate}
\item If we set $n_r = 2$, the diversity performance in
  \eqref{eq:nt2_firstsegment}  is equal to $2n_t - \lp n_t+1\rp r$;
  this overlaps with the  optimal tradeoff curve of the channel
for small enough multiplexing gains, i.e., $ r \leq 1$, thus achieving
the
first segment for the  $n_t \times 2$ i.i.d.\ Rayleigh fading channel
(see Figure \ref{fig:dblast_22}).

\begin{figure}[h]
\begin{center}
\scalebox{0.8}{\input{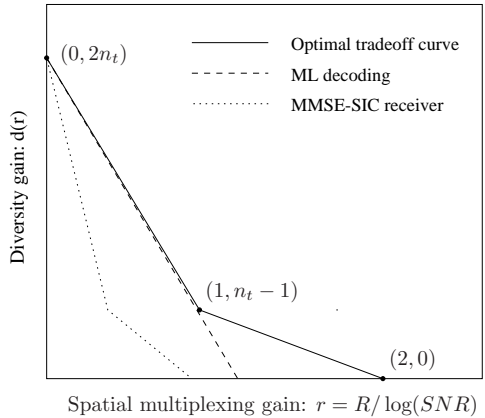}} \caption{Diversity
performance of the D-BLAST architecture.} \label{fig:dblast_22}
\end{center}
\end{figure}
\item From the perspective of one of the streams in the D-BLAST
  architecture, the best diversity performance is obtained if the
  other stream didn't exist at all (or was decoded correctly and thus
  canceled exactly). Suppose this is the case: then each data stream
  sees a parallel channel with $n_t$ scalar sub-channels, each of
  whose squared amplitudes are i.i.d.\ with distribution
  $\chi^2_{2n_r}$. The optimal tradeoff curve for this parallel
  channel with a data rate of $\lp n_t+1\rp/2$ bits/symbol (cf.\
  \eqref{eq:singlestreamrate}) is
\beq\label{eq:dgain_upperbound}
    n_r\lp n_t  - \frac{n_t + 1}{2} r\rp.
\eeq
The diversity performance of any data stream with the other stream
  being perfectly canceled cannot be any more than the gain in
  \eqref{eq:dgain_upperbound}. However, from the claim in
  Proposition~\ref{prop:dblast_firstsegment_rayleigh} (cf.\
  \eqref{eq:nt2_firstsegment}), we   observe that this upper
  bound is {\em exactly} equal to the diversity gain achieved even
  when there is inter-stream interference. There we conclude:
\begin{quote} {\em Under the
  joint ML decoder, inter-stream  interference  is not the typical
  error event}.
\end{quote}
 We study the joint ML decoder  in some detail in  the
  next section.
\item Finally, we observe that we crucially used the symmetry between
  the two streams in the above argument. With more than two streams,
   the middle streams see more interference than the outer two streams
  and an extension to this situation is not natural.
\end{enumerate}

\subsubsection{D-BLAST and ML Decoding}\label{sec:dblast_ml}

In this section, we discuss the ML decoding of the two data streams in
the D-BLAST architecture in some detail. To make our discussions
simple and concrete we focus on the simple case of $n_t = 2$; the
received signal spans three time symbols and can be written as
\begin{eqnarray}
\nno [\by_1 \by_2 \by_3] & = & \left[ \bh_1 \bh_2 \right] \Lbr
\begin{array}{ccc}
0  & p_2 & q_2  \\
p_1 & q_1 & 0
\end{array} \Rbr + \Lbr \vw_1 \vw_2 \vw_3\Rbr.
\end{eqnarray}
The two data streams $\Lbr p_1, p_2\Rbr$ and $\Lbr q_1,q_2\Rbr$ are
unit block-length  approximately universal codes for a parallel
channel with 2 sub-channels and independent of each other. For
concreteness, suppose $p_1$ ($q_1$) and $p_2$ ($q_2$) are points from
a  QAM constellation and correspond to bit reversal with alternative
bits flipped of each other (cf.\
Section~\ref{sec:altenatebitflip}). The ML
decoder  makes a joint decision on both these codes using the three
received vectors $\vy_1,\vy_2,\vy_3$.  However, due to the specific
structure of the zeros in the D-BLAST architecture, the joint ML
decoder  can be broken down algorithmically into three separate steps:
\begin{enumerate}
\item We observe that the received vector at the first time symbol
  $\vy_1$ gives information only about the  the QAM symbol $p_1$:
\beq
 \by_1  =  p_1 \bh_2 + \vw_1.
\eeq
In particular, $\by_1$ specifies exactly the most significant bits of
  the bit representation of the QAM point $p_1$ (cf.\
  Section~\ref{sec:explicit}). More specifically, the number of MSBs
  of $p_1$ that can be deduced from $\vy_1$ is with high probability
  equal to    $\lfloor\log\lp|\bh_2|^2\SNR\rp\rfloor$; further more,
  the information about the remaining bits of $p_1$ depends on the
  noise $\vw_1$ that is independent of the received signals at the
  other two time symbols. Since the QAM  points $p_1$
  and $p_2$ correspond to bit reversals (with alternate bits flipped)
  of each other, we have also deduced the {\em least} significant bits
  of $\lfloor\log\lp|\bh_2|^2\SNR\rp\rfloor$ of $p_2$.
\item The scenario at the third time symbol is identical to that at
  the first time symbol except that $p_1$ is replaced by $q_2$ and
  $p_2$ by $q_1$. In particular, we can deduce
  $\lfloor\log\lp|\bh_1|^2\SNR\rp\rfloor$ MSBs of $q_2$ (and the
  $\lfloor\log\lp|\bh_1|^2\SNR\rp\rfloor$ LSBs of $q_1$) from
  $\vy_3$; further more, the information about the remaining bits of
  $q_2$ (and hence $q_1$) depends on the noise vector $\vw_3$ that is
  independent of  the received vector at the first two time symbols.
\item We are now ready to focus on the received vector at the second
  time  symbol:
\beq\label{eq:reduce2vblast}
 \vy_2 = p_2 \vh_1 + q_1 \vh_2 + \vw_2.
\eeq
Here we know some of the LSBs of both $p_2$ and $q_1$ (due to
  processing of the  received vector at the first and third time
  symbols, respectively); this reduces the randomness in $p_2$ and
 $q_1$ to another sparser QAM which is a subset of the original QAM
  from which they were drawn. We see from \eqref{eq:reduce2vblast} is
  exactly the output of a $2\times 2$ MIMO channel with uncoded QAMs
  transmitted over the two transmit antennas, i.e., uncoded QAM
  transmission over the V-BLAST architecture. Thus, the ML decoding of
  the two streams of the D-BLAST architecture reduces to that of a
  decoding uncoded QAM transmission over the V-BLAST architecture.
 \end{enumerate}

\subsubsection{A Time-Space Code and the Second Segment}
 While we have seen the tradeoff optimality of the D-BLAST
 architecture in achieving the first segment of the $n_t\times 2$
 i.i.d.\ Rayleigh fading channel, there is a simple transformation of
 this architecture that achieves the {\em second segment} of the same
 channel. The key is to consider a {\em time-space} version of the
 space-time D-BLAST  architecture: replace the transmit symbol at time
 symbol $m$ over the  transmit  antenna $k$ by the transmit symbol at
 time symbol $k$ and transmit antenna $m$. In particular, the
 time-space version of the space-time code in \eqref{eq:dblastnt2} is
\beq\label{eq:dblastnt22}
\Lbr \begin{array}{cccc}
0  & \cdots & 0 & p_1  \\
\vdots & \adots & p_2 & q_1 \\
0 & \adots & q_2 & 0 \\
p_{n_t} & \adots & \adots & \vdots \\
q_{n_t} & 0 & \cdots & 0
\end{array} \Rbr.
\eeq It is meant to be used over a channel with $n_t+1$ transmit
antennas and spans $n_t$ time symbols long; observe that the
original code in \eqref{eq:dblastnt2}  is meant to be used over a
channel with $n_t$ transmit antennas and spans $n_t+1$ time symbols
long. Suppose that $\Lbr p_1, \ldots ,p_{n_t}\Rbr$ and $\Lbr
q_1,\ldots ,q_{n_t}\Rbr$ independent unit block-length approximately
universal codes for the parallel channel at rate $0.5 \log\SNR$ bits
per sub-channel; this corresponds to the overall code in
\eqref{eq:dblastnt22} to have a  total multiplexing  rate of $r$
bits/symbol. Our main result is a precise characterization of the
diversity performance of this space-time code over the i.i.d.\
Rayleigh fading channel; the proof is available in
Appendix~\ref{ap:timespace}.
\begin{prop}\label{prop:timespacerayleigh}
The diversity gain of joint ML decoding  the data streams of the
time-space code in \eqref{eq:dblastnt22} at a total multiplexing
rate of $r$ bits/symbol over the $(n_t+1)\times n_r$ i.i.d.\
Rayleigh fading MIMO channel with $n_r \geq 2$ is equal to \beq
\label{eq:timespacediversity}
 \frac{n_t n_r}{2}(2 -  r). \eeq
\end{prop}

\begin{figure}[h]
\begin{center}
\input{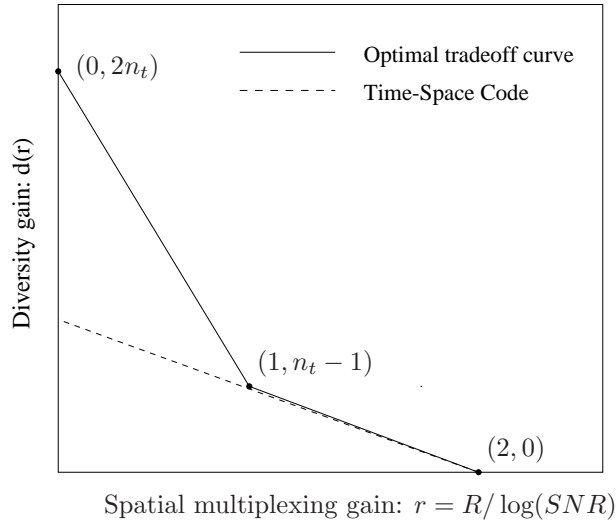}
\caption{D-BLAST curves vs the optimal tradeoff
curve}\label{fig:dblast32}
\end{center}
\end{figure}

Setting $n_r = 2$, we see that the diversity gain  in
\eqref{eq:timespacediversity} is equal to $2n_t - n_tr$ which overlaps
with the optimal tradeoff curve for that channel for large enough
multiplexing gains, i.e., $r \geq 1$; in particular, this achieves the
second segment of the tradeoff curve (see Figure~\ref{fig:dblast32}).

\subsubsection{Tradeoff Optimality over Isotropic
  Channels}\label{sec:dblast_iso} We demonstrate the robustness of
the performance results  a time-space code for the i.i.d.\ Rayleigh
fading channel by generalizing them to the class of isotropic fading
distributions: in particular, we are interested in MIMO channel
distributions which satisfy the property in
\eqref{eq:isotropic_definition}. Further recall the definition of
the polynomial decay rates of the squared singular values of the
$n\times n$ MIMO channel in \eqref{eq:decayrate_singularvalues}. The
proofs of the results in this section are available in
Appendix~\ref{ap:mimo_isotropic}.

%Our first result is the restricted approximate universality
%performance of the D-BLAST architecture with two data streams in
%achieving the first segment of the tradeoff curve; this generalizes
%the result in Proposition~\ref{prop:dblast_firstsegment_rayleigh}.

%\begin{thm}
% The D-BLAST architecture in \eqref{eq:dblastnt2} with approximately
%universal parallel channel codes as its two data streams operated at a
%total multiplexing gain of $r$ over any isotropic $n_t\times 2$ MIMO
%channel achieves the first segment of its tradeoff curve, provided
%\begin{eqnarray*}
%                 k_1   \leq   n_t -2  & \mbox{and} & 0  \leq  k_2
%                -k_1  \leq  2.
%\end{eqnarray*}
%\end{thm}

Our result is the restricted approximate universality of the
time-space version of the D-BLAST architecture with two data streams
in achieving the second segment of the tradeoff curve; this
generalizes the result in Proposition~\ref{prop:timespacerayleigh}.
The proof of this result is available in
Appendix~\ref{ap:mimo_isotropic}.

\begin{thm}
The  diversity gain of joint ML decoding  the data streams of the
time-space code in \eqref{eq:dblastnt22} at a total multiplexing rate
of $r$ bits/symbol over any isotropic $n_{t+1} \times 2$ MIMO channel
achieves the second segment of its tradeoff curve, provided
\begin{eqnarray*}
                 k_2 -k_1 & \geq & 2,
\\\nno k_1 & \leq & 0.
\end{eqnarray*}
\end{thm}

\section{Conclusion}
We have presented a precise characterization of universally-tradeoff
optimal codes for the MIMO channel. We also presented explicit codes
for the parallel channel that are simple to encode and decode. These
codes, along with the general construction in \cite{VG05},
completely solves the code design problem for the parallel channel.
For the MIMO channel, we suggest using the D-BLAST architecture to
reduce it to a parallel channel and using codes designed for the
parallel channel.  This approach is reasonable when the block-length
is large, since in  this case the initialization overhead in D-BLAST
is  insignificant. While, finite block length approximately universal
codes for the MIMO channel have been constructed, they are not known
to be simple to decode; construction of  simple codes for the MIMO
channel remains an open problem.

   Alternative to approximately universal codes for MIMO channel, we
have seen the existence of simple codes for the MIMO channel that are
approximately universal for a restricted class of fading
distributions. Our construction has been restricted for specific
number of antenna elements; a  generalization of this construction is
also an interesting future research direction.

\appendix

\section{Converse for Approximate Universality}\label{ap:precise_char}

We want to show that if a coding scheme does not satisfy the
universal code design criterion, then there exists a fading
distribution such that the coding scheme is not tradeoff optimal. In
the high SNR scaling of \cite{ZT03}, a coding scheme is defined by a
{\em discrete sequence} of codes $C(\SNR)$ with rate $r\log\SNR$. If
this sequence does not satisfy the approximate universality
criterion, then there exists a subsequence of $C(\SNR)$ such that
for every code in the sub-sequence there exists a codeword pair such
that it does not satisfy the universal criterion. For proving the
existence of a fading distribution such that the original sequence
is not tradeoff optimal, it is enough to find a fading distribution
for which this subsequence of codes is not tradeoff-optimal.
Therefore we assume that for every code in the sequence we can find
a codeword pair that does not satisfy the universal criterion.

A brief note regarding our notation: we use the symbols $\doteq (
\dotgq, \dotlq)$ to denote exponential equality (inequality), i.e.,
\begin{eqnarray}
\nno f(\SNR) \doteq \SNR^b & \impls & \lim_{\SNR \tends \infty}
\frac{\log f(\SNR)}{\log \SNR} = b.
\end{eqnarray}

\subsection{Proof of Theorem~\ref{thm:sub}}
\label{ap:parallel_mimo} Here we focus on the necessity of the
condition for approximate universality for the MIMO channel. If a
sequence of codes is not approximately universal, we show that there
exists an i.i.d.\ distribution on $\psi_\ell$s such that this
sequence of codes is not tradeoff optimal.

For codewords $\mX_A$ and $\mX_B$, the pairwise error conditioned on
a channel realization, $\mH$, can be written as (cf.\
\rref{eq:prwse_pe}): \beq\nonumber \mathbb{P}_e\lp \mX_A \rightarrow
\mX_B| \mH \rp = Q\lp\sqrt{\frac{\SNR\sum_{\ell=1}^{\nmin}
|\lambda_{\ell}|^2|\psi_{\ell}|^2}{2}}\rp. \eeq  The approximate
universality condition can then be written as: \bqa\nno
\min\limits_{\sum_{\ell=1}^{\nmin}\log\lp1+|\psi_{\ell}|^2\SNR\rp
> r\log\SNR}\SNR\sum_{\ell=1}^{\nmin} |\lambda_{\ell}|^2|\psi_{\ell}|^2 &
\dotgq & 1. \eqa Thus, if a sequence of codes does not satisfy the
universal criterion then there exists a sequence of codeword pair
differences, $\mD(\SNR)$, and a corresponding realization
$\mH^a(\SNR)$ such that \beq\label{eq:d_ha}
\SNR\sum_{\ell=1}^{\nmin}
|\lambda_{\ell}(\SNR)|^2|\psi_{\ell}^a(\SNR)|^2 < 2^{-r\eps\log\SNR
} \eeq for some positive $\eps$, where $\mH^a(\SNR)$ satisfies
\beq\label{eq:outage_a}\sum_{\ell=1}^{\nmin}\log\lp1+|\psi_{\ell}^a(\SNR)|^2\SNR\rp
= r\log\SNR. \eeq Now define $\mH^b(\SNR)$ as \beq\nno
|\psi_{\ell}^b(\SNR)|^2 = |\psi_{\ell}^a(\SNR)|^2\cdot
2^{r\eps\log\SNR} , \quad l=1,\ldots,L.\eeq Then using
\rref{eq:d_ha} and \rref{eq:outage_a}, $\mH^b(\SNR)$ satisfies
\bqa\label{eq:pe_hb} \SNR\sum_{\ell=1}^{L}
|\lambda_{\ell}(\SNR)|^2|\psi_{\ell}^b(\SNR)|^2 & < & 1 \quad {\rm
and}
\\\label{eq:hepsoutage}
\sum_{\ell=1}^L\log\lp1+|\psi_{\ell}^b(\SNR)|^2\SNR\rp & \geq &
r(1+\eps)\log\SNR. \eqa
%Thus, for every value of $\SNR$ (in the sequence) we associate a
%channel realization, $\mH^b(\SNR)$, such that it is not in $\CH_{\rm
%out}(r(1 + \eps/2), \SNR)$ and there exists a pair of codewords in
%$C(\SNR)$ such that \beq\nonumber \mathbb{P}_e\lp \bx_A \rightarrow
%\bx_B|\mH^b(\SNR)\rp \geq   Q(1). \eeq
Now, consider the i.i.d.\ fading distribution on $\mH$ such that:
\bqa\label{eq:iid_h} \prob{|\psi_{\ell}|^2 \leq \frac{1}{x}} &
\doteq & \frac{1}{x^{\frac{2}{\eps}}}\quad\forall\quad
\ell=1,\cdots,L. \eqa The diversity for the code-sequence can then
be upper bounded using the following sequence of steps:
\begin{enumerate}
\item The pairwise error for the codeword difference $D(\SNR)$
can be lower bounded by a constant, $Q(\sqrt{0.5})$, for a range of
channels such that (see \rref{eq:pe_hb}): \bqa\nno \left\{ \mH:
|\psi_{\ell}|^2 < |\psi^{b}_\ell(\SNR)|^2,\quad\ell=1,\cdots,L
\right\}.\eqa Furthermore, because of the power constraint on the
input we can assume that \bqa\label{eq:honebysnr}
|\psi_{\ell}^b(\SNR)|^2 & \dotgq & \frac{1}{\SNR} \quad\forall\quad
\ell=1,\cdots,\nmin \eqa If this is not true, we can increase
$\psi_{\ell}^b(\SNR)$ to $\frac{1}{\nmin\SNR}$ such that
\rref{eq:hepsoutage} and \rref{eq:pe_hb} still hold.

\item  Hence the probability of error can be lower
bounded by \bqa\label{eq:pe_Hstar2} \mathbb{P}_e(C(\SNR)) & \geq &
\frac{Q(\sqrt{0.5})\prod\limits_{\ell=1}^{\nmin}\prob{|\psi_{\ell}|^2
\leq |\psi_{\ell}^b(\SNR)|^2}}{\SNR^r}. \eqa Writing \bqa\nno
|\psi_{\ell}^b(\SNR)|^2 = \SNR^{-\alpha_\ell},\eqa the probability
of error expression \rref{eq:pe_Hstar2} can be written as (also see
\rref{eq:iid_h}): \bqa\label{eq:pe_alpha} \mathbb{P}_e(C(\SNR)) &
\dotgq & \SNR^{-\lp\frac{2}{\eps}\sum_{\ell}\alpha_\ell+r\rp}, \eqa
where $\alpha_\ell$s satisfy (see \rref{eq:hepsoutage} and
\rref{eq:honebysnr}): \bqa\nno \sum_{\ell} (1-\alpha_\ell)^+ \geq
r(1+\eps) &{\rm and}& \alpha_\ell  \leq  1. \eqa Therefore \bqa
\nno\sum_{\ell}\alpha_\ell & \leq & \nmin-r(1+\eps). \eqa Then the
probability of error for $C(\SNR)$ is lower bounded by (see
\rref{eq:iid_h} and \rref{eq:pe_alpha}): \bqa\nno
\mathbb{P}_e(C(\SNR)) & \dotgq &
\SNR^{-\lp\frac{2}{\eps}(\nmin-r(1+\eps))+r\rp},
\\\nno & = & \SNR^{-\lp\frac{2}{\eps}(\nmin-r)-r\rp}.
\eqa
\end{enumerate}
Thus, the diversity of the sequence of codes is upper bounded by:
\bqa\label{eq:pe_parallel_lower} \frac{2}{\eps}(\nmin-r) - r. \eqa
The outage curve on the other hand is given by\footnote{A proof for
this result can be seen from Appendix~\ref{ap:miso}, Equation
\rref{eq:par_miso_outage}, with $n_t$ replaced by $\nmin$ and $n_tr$
replaced by $r$ and $a$ replaced by $\frac{2}{\eps}$.}: \bqa\nno
\frac{2}{\eps}\lp \nmin - r \rp. \eqa Thus, comparing with
\rref{eq:pe_parallel_lower}, this sequence of codes is not tradeoff
optimal and hence not approximately universal.

\section{Proof of Theorem
  \ref{thm:random_permutations}}\label{ap:random_permutations}
Consider a parallel slow fading channel with $L$ sub-channels.
A permutation code over this channel can be rewritten as
\begin{eqnarray}
\nno \Co = \lbr \sqrt{\frac{\SNR}{\SNR^r}}\lp
q,f_2(q),...,f_{L}(q)\rp| q \in \zQ\rbr,
\end{eqnarray}
where \beq \zQ = \lbr\lp a+ib\rp : -\frac{\SNR^{r/2}}{2} \leq a,b
\leq \frac{\SNR^{r/2}}{2} \rbr\eeq is the integer-QAM with $\SNR^r$
points, and $f_2,...,f_{L}$ are  permutations of $\zQ$. We
define the normalized product distance between two codewords as
\begin{eqnarray}\label{eq:define_pid}
\pi_d\lp q_1, q_2 \rp & = & \frac{|q_1-q_2|^2}{\SNR^r}
\prod_{k=2}^{L} \frac{|f_k(q_1) - f_k(q_2)|^2}{\SNR^r}.
\end{eqnarray}
The condition for approximate universality, \rref{eq:par_univ_crit},
on the other hand, can be written as
\begin{eqnarray}\label{eq:pidist}
\pi_d \lp q_1, q_2 \rp & \dotgq & \frac{1}{\SNR^r} \qquad \forall
 q_1 \not= q_2.
\end{eqnarray}

The number of permutation codes with $\SNR^r$ points
is given by
\begin{eqnarray}
\nno ((\SNR^r)!)^{L-1}.
\end{eqnarray}
We now prove existence of a permutation code in this ensemble such
that \rref{eq:pidist} is satisfied. We average of the {\em inverse}
of product distance over all such codes under the uniform measure
(all codes have the same probability). The intuition behind
averaging the inverse of product distance is to capture the codeword
differences that have small product distance, which is the event of
interest.

\begin{eqnarray}\nno
\E \Lbr \frac{1}{\pi_d} \Rbr & = & \frac{1}{(\SNR^r!)^{L-1}
\SNR^{2r}} \sum_{\substack{f_2,...,f_{L},\\q_1, q_2 \neq q_1}}
\frac{\SNR^r}{|q_1-q_2|^2} \prod_{k=2}^{L}\frac{\SNR^r}{ |f_k(q_1) -
f_k(q_2)|^2}, \\ \nno
 & = & \frac{((\SNR^r-2)!)^{L-1}\SNR^{rL} }{(\SNR^r!)^{L-1}
\SNR^{2r}} \sum_{\substack{q_1, q_2 \neq q_1,\\ q_1^{k}, q_2^k \neq q_1^{k} \\
k = 2,\cdots,L}} \frac{1}{|q_1-q_2|^2} \prod_{k=2}^{L}\frac{1}{
|q_1^k - q_2^k |^2}.
\end{eqnarray}
The second equality is obtained by considering all permutations $f_k$'s
that map $q_1$ to $q_1^k$ and $q_2$ to $q_2^k$; the number of such
permutations is $((\SNR^r-2)!)^{L-1}$. Therefore,
\begin{eqnarray}
\nno E\Lbr\frac{1}{\pi_d}\Rbr & \leq & \frac{1}{\SNR^{Lr}} \Lbr
\sum_{q_1, q_2\neq q_1} \frac{1}{|q_1-q_2|^2} \Rbr^{L}.
\end{eqnarray}
Because of the symmetry of the QAM, the average inverse product
distance  can be  further upper bounded as
\begin{eqnarray}
\label{eq:sumpar} \E\Lbr\frac{1}{\pi_d}\Rbr & \leq & \frac{1}{\SNR^{Lr}} \Lbr
\SNR^r \sum_{q_1 \neq 0} \frac{1}{|q_1|^2} \Rbr^{L}, \\\nno
 & = &  \Lbr \sum_{q_1 \neq 0} \frac{1}{|q_1|^2} \Rbr^{L}.
\end{eqnarray}

The summation inside the parantheses in \eqref{eq:sumpar} can be
upper bounded by $\lp\log\SNR\rp^2$. This implies that  the
expectation can be upper bounded by
\begin{eqnarray}
\nno \E\Lbr\frac{1}{\pi_d}\Rbr & \dotlq &  1.
\end{eqnarray}
We conclude that  there exists at least one permutation code $\Co^a$
with
the average inverse product distance less than $1$. We now use this
code $\Co^a$ with good average behavior to construct a code that has a
good worst-case behavior. For $\Co^a$, \bq \nno
\frac{1}{\SNR^{2r}}\sum_{q_1 \neq q_2 \in \zQ}
\frac{1}{\pi_d(q_1,q_2)} & \dotlq & 1. \eq Therefore, \bqa\nno
\frac{1}{\SNR^{r}}\sum_{q_1 \in \zQ} g(q_1) & \dotlq & \SNR^r, \eqa
where \bqa\nno g(q_1) & = & \sum_{q_2 \in \zQ, q_2 \neq q_1}
\frac{1}{\pi_d(q_1,q_2)}. \eqa Thus, at least half of the $q_1$'s
have $g(q_1)   \dotlq   \SNR^r$. By expurgating at most half
the codewords, we can construct a code $\Co^b$ such that:
\bqa\label{eq:gq_ub} g(q_1) & \dotlq & \SNR^r  \qquad \forall
q_1. \eqa This implies that for every $q_2\not= q_1$, \beq\nonumber
\pi_d\lp q_1, q_2 \rp   \dotgq   \frac{1}{\SNR^r}; \eeq this is
precisely the criterion for approximate universality  \rref{eq:pidist}.
Finally, expurgating at most of half of the codeword reduces the rate
of the code  by
at most one and hence does no change the multiplexing gain. Thus,
there exist approximately universal permutation codes.

\subsection{Product distance distribution}\label{ap:pdd}

A  statement much more stronger that that  made about the code
$\Co^b$ constructed in Section~\ref{ap:random_permutations}. The
result below  characterizes the behavior of the product distance
$\pi_d$, cf.\ \eqref{eq:define_pid}, (rather than just a lower
bound, which is what was required for
approximate universality), and hence can be thought of as a weight
distribution result for the product distance.

\begin{thm}
Consider a parallel slow fading channel with $L$ sub-channels.
There exists a permutation code  with $\SNR^r$
points over this channel such that the number of codeword pairs that
have a product
distance less than $\SNR^{k-r}$ is $\Theta(\SNR^{r+k})$, for $k$ in
$\left[ 0,r \right]$.
\end{thm}
\begin{proof}

We start with the code $\Co^b$ constructed above that satisfies
\rref{eq:gq_ub}: then for each $q_1$, the number of codewords which
are at a product distance less than $\SNR^{k-r}$ is
$\Theta(\SNR^{k})$, for $k$ in $\left[ 0,r \right]$ (otherwise such
a code will not satisfy \rref{eq:gq_ub}). Considering all possible
values of $q_1$, the number of codeword difference that have product
distance less than $\SNR^{k-r}$ is $\Theta(\SNR^{r+k})$, for $k$ in
$\left[ 0,r \right]$.

\end{proof}

\section{Proof of Theorem \ref{thm:abf}}\label{ap:abf}

Let the binary representation of integers $a_1$ and $a_2$ be:
\begin{eqnarray}\nno
a_1 & = & b_{R/2}^1 \cdots b_1^1,
\\\nonumber a_2 & = & b_{R/2}^2 \cdots b_1^2.
\end{eqnarray}
Let $k$ be the largest integer such that $b_k^1 \not= b_k^2$. Then
without any loss of generality we can assume that $b_k^1 = 1$ and
$b_k^2 = 0$. We also write
\begin{eqnarray}\nno
b_i = b_i^1 = b_i^2 & &\forall \quad k+1 \leq i \leq {R/2},
\end{eqnarray}
for notational convenience as well as to emphasize that the largest
${R/2}-k$ bits are the same. Now, let $l$ be the smallest integer
such that $b_{k-l}^1 \geq b_{k-l}^2$. Note that this implies that
\begin{eqnarray*}
b_{k-i}^1 = 0 & & b_{k-i}^2 =1 \quad \forall \quad 1\leq i \leq l-1,
\end{eqnarray*}
which is similar to the codeword pair that was the counter example
given for the fact that simple bit-reversal is not universal (see
Section \ref{sec:prodbit}). Here we essentially prove that such
pairs are the only reason that the simple bit reversal is not
approximately universal and bit reversal with alternated bit
flipping can tackle this problem. We consider the following
subcases:

\begin{itemize}

\item  If no such $l$ exists, Then $a_i$s can be written as:
\begin{eqnarray*}
a_1 & = & b_{R/2}\cdots b_{k+1} 1 0 \cdots 0
\\ a_2 & = & b_{R/2}\cdots b_{k+1} 0 1 \cdots 1
\end{eqnarray*}
and $B(a_i)$s can be written as:
\begin{eqnarray*}
B(a_1) & = & 101 \cdots \\
B(a_2) & = & 010 \cdots
\end{eqnarray*}
Thus $B(a_1)-B(a_2)$ is lower bounded by $2^{{R/2}-2}$, hence
\rref{eq:abfpd} is satisfied.

\item $l \geq 2$: Then $a_i$s  can be written as:
\begin{eqnarray*}
a_1 & = & b_{R/2}\cdots b_{k+1} 1 0 \cdots 0 b_{k-l}^1
b_{k-l-1}^1\cdots
b_1^1 \\
a_2 & = & b_{R/2}\cdots b_{k+1} 0 1 \cdots 1
b_{k-l}^2b_{k-l-1}^2\cdots b_1^2
\end{eqnarray*}
then, the difference $a_1-a_2$ can be lower bounded by $2^{k-l-1}$
and $B(a_i)$s can be written as:
\begin{eqnarray*}
 B(a_1) & = & \overline{b_1^1}b_2^1 \cdots \overline{b_{k-l-1}^1}
b_{k-l}^11101 \cdots 0b_{k+1}\cdots b_{R/2} \\
B(a_2) & = & \overline{b_1^1} b_2^2 \cdots
\overline{b_{k-l-1}^2}b_{k-l}^2 0010 \cdots 1b_{k+1}\cdots b_{R/2}
\end{eqnarray*}

Then the difference $|B(a_1)-B(a_2)|$ is lower bounded by
$2^{{R/2}-(k-l)-2}$ (here we have assumed that $b_{k-l}$ is not
flipped, \ie $k-l$ is even; if $k-l$ is odd, then same argument hold
with $a_2$ and $a_1$ reversed). Thus, the product distance is lower
bounded by $\frac{1}{8\cdot2^{{R/2}}}$ (which is the one in
\rref{eq:abfpd}).

\item If $l = 1$: then $a_i$s can be written as:
\begin{eqnarray*}
a_1 & = & b_{R/2}\cdots b_{k+1}1 b_{k-1}^1\cdots
b_1^1 \\
a_2 & = & b_{R/2}\cdots b_{k+1}0 b_{k-1}^2 \cdots b_1^2
\end{eqnarray*} then the difference $a_1-a_2$ can be lower bounded
by $2^{k-2}$ (since $b_{k-1}^1 \geq b_{k-1}^2$). The $B(a_i)$s can
be written as:
\begin{eqnarray*}
B(a_1) & = & \overline{b_1^1}b_2^1 \cdots \overline{b_{k-1}^1}
1\overline{b_{k+1}}b_{k+2}\cdots b_{R/2} \\
B(a_2) & = & \overline{b_1^2} b_2^1 \cdots \overline{b_{k-1}^2}
0\overline{b_{k+1}}b_{k+2}\cdots b_{R/2}
\end{eqnarray*}

and the difference $|B(a_1)-B(a_2)|$ is lower bounded by
$2^{{R/2}-k}$ (here we have assumed that $b_{k-1}$ is flipped, \ie
$k$ is even; same is true if $k$ is odd). Thus, the product distance
is lower bounded by $\frac{1}{4\cdot2^{{R/2}}}$.

\end{itemize}

\section{Proof of Theorem \ref{thm:universal_decoding}}\label{ap:universal_decoding}

We again consider the I and Q channels separately. Then we want to
define $L-1$ permutations of the PAM such that the corresponding
permutation code is approximately universal. We consider the
$q$-digit representation of the PAM. For a PAM with $q^n$ points and
number it from left to right by $0$ to $q^n-1$ (in term of the rate
$R$, $n$ behaves like $\frac{\log_2{R/2}}{log_2q}$). For showing
that a universally decodable system satisfies the product distance
criterion, we have to resort to irregularly spaced PAMs. For every
$m\th$ least significant q-bit change, we put a gap of $g q^{m-1}$.
Similar to the two sub-channel case, using this construction we
prove that any universally decodable scheme satisfies the condition
for approximate universality: consider any two codewords; suppose
for the $\ell\th$ sub-channel their $k_\ell$ MSBs are the same and
$(k_\ell+1)\th$ MSB is different. By construction of  the
irregularly spaced QAM, the normalized (by $q^n$) separation in  the
$\ell\th$ coordinate is lower bounded by \beq\nonumber
\frac{gq^{n-k_\ell-1}}{q^n} = gq^{-(k_\ell+1)}. \eeq The universal
decodability condition implies that  if $\sum_{\ell} k_\ell \geq n$,
then there exists a unique codeword corresponding to the MSBs.
Therefore, the $k_\ell$s must satisfy \beq\nonumber \sum_{\ell}
k_\ell < n.\eeq Thus, the product distance can be lower bounded by
\begin{eqnarray}
\nno |d_1d_2\cdots d_L|^{2/L} & \geq & \lp\prod_{\ell} g^2
q^{-2k_\ell-2} \rp^{1/L}, \\\nno & = &   \lp q^{-2\sum_\ell
k_\ell}\rp^{1/L}, \\\nno & \geq & \frac{g^2}{q^2} q^{-2n/L}
\\\label{eq:prod_d_q} & = & \frac{g^2}{q^22^{R}}
\\\nno & \doteq & \frac{1}{2^R},
\end{eqnarray}
implying that the code satisfies the approximate universality
condition \rref{eq:par_univ_crit}. For a PAM of size $q^n$, the
(normalized) increase in size is given by
\begin{eqnarray}\nno
\sum_{m=1}^n g q^{m-1-n} (\mbox{number of $m\th$ LS q-bit changes})
& = & \sum_{m=1}^n g q^{m-n} (q^{n-m}),
\\\label{eq:gain_q} &=& g n.
\end{eqnarray}
In the high SNR scaling,
\begin{eqnarray*}
q^n =  \SNR^r & \implies & gn  \doteq  \log\SNR.
\end{eqnarray*}
Thus the extra spacing does not affect the multiplexing gain.

We also note the Theorem \ref{thm:universal_decoding} is true even
if the field size $q$ is growing like $\log\SNR$. Note that if $q$
grew like a polynomial in $\SNR$, \ie like $\SNR^{\eps}$, then we
can no longer ignore $q$ in \rref{eq:prod_d_q} and such a code then
will not be approximately universal. We also have to show that the
power gain because of the gaps still increases slowly enough so as
to not affect the multiplexing gain. For a PAM of size $q^n$, the
increase in size is, cf.\ \rref{eq:gain_q},
\begin{eqnarray*}
gn & \dotlq & \SNR^{\eps} \quad \forall \quad \eps >0.
\end{eqnarray*}
Therefore, the extra spacing does not affect the diversity-multiplex
tradeoff.

\section{Proof of Propositions~\ref{prop:miso_2}}\label{ap:miso}
We use an approximately universal parallel channel code, (e.g.\ a
permutation code $[p_1,\cdots,p_{n_t}]$ with total rate $Rn_t$) over
the MISO channel in a diagonal fashion:
\begin{equation}\label{eq:miso-one}
\left[ \begin{array}{ccc} p_1 & 0 & 0 \\
0 & \ddots & 0 \\
0 & 0 & p_{n_t} \\
\end{array}\right].
\end{equation}
We prove that scheme \rref{eq:miso-one} is tradeoff optimal for MISO
channel with i.i.d.\ fading coefficients. Since it operationally
converts the MISO channel into a parallel channel, we only need to
match the outage probabilities of the MISO channel and the
corresponding parallel channel. The outage probability of the MISO
channel is given by.
\begin{eqnarray}\label{eq:MISO-outage}
\prob{\log\lp1+\sum_{i=1}^{n_t}|h_i|^2\SNR\rp \leq
r\log\SNR}
\end{eqnarray}
For the equivalent parallel channel, the outage probability is given
by
\begin{eqnarray}\label{eq:miso-parallel-outage}
\prob{\sum_{i=1}^{n_t} \log \lp 1+|h_i|^2\SNR\rp \leq
n_tr\log\SNR}
\end{eqnarray}
The near zero behavior of sum of $|h_i|^2$s can be upper and lower
bounded as:
\begin{equation*}
\lp\prob{|h_1|^2 < \frac{x}{n_t}}\rp^{n_t} \leq
\prob{\sum_{i=1}^{n_t}|h_i|^2 \leq x} \leq \lp\prob{|h_i|^2 \leq
x}\rp^{n_t}.
\end{equation*}
Since the upper and lower bound have the same decay rate, the
probability of outage of the MISO channel, \rref{eq:MISO-outage},
has a decay rate of
\begin{equation}\label{eq:outage_miso}
\lp\frac{\SNR^r}{\SNR}\rp^{an_t}.
\end{equation}
Thus, the outage curve of the MISO channel with i.i.d.\ fading
coefficients with the $a$ denoting the decay rate of $|h_1|^2$ near
zero is \beq\nno d_{\rm out}(r) = an_t(1-r). \eeq

The second outage probability, \rref{eq:miso-parallel-outage}, is
somewhat more involved. Define $\alpha_i$ by  \begin{equation*} |h_i|^2
 = \frac{\SNR^{\alpha_i}}{\SNR}. \end{equation*} In this notation,
the outage condition for the parallel channel can be written as
\begin{equation}\label{eq:outage_par_alpha}
\sum_i\alpha_i   \leq   n_tr.
\end{equation}
Since the sub-channels are independent, the outage probability (cf.\
\rref{eq:miso-parallel-outage}) has the decay rate
\begin{equation}\label{eq:outage_opt}
\max_{\alpha_1,\cdots,\alpha_{n_t}} \prod_{i=1}^{n_t}\lp
\frac{\SNR^{\alpha_i}}{\SNR}\rp^a,
\end{equation}
where the maximization is under the constraint in
\rref{eq:outage_par_alpha}. Thus, the decay rate of the outage
probability  expression in
\rref{eq:outage_opt} is
\begin{equation}\label{eq:par_miso_outage}
\lp\frac{\SNR^r}{\SNR}\rp^{n_ta},
\end{equation}
the same as that in \rref{eq:outage_miso}; this  completes the proof.
%Note that here we don't need $h_i$s to be identically distributed.
%We only need independence of $h_i$s and a similar near zero behavior
%(same $a$ for all $i$).

\section{Proof of Proposition \ref{prop:dblast_firstsegment_rayleigh}}\label{ap:dblast}
We prove that the diversity obtained by the code in
\rref{eq:dblastnt2} is $n_r\lp n_t- \tilde{r}\rp$, where
$\tilde{r}\log\SNR$ is the rate of codes $[p_1,\cdots,p_{n_t}]$ and
$[q_1,\cdots,q_{n_t}]$.

The pairwise probability of error, averaged over the Rayleigh fading
channel with $n_r$ receive antennas is given by \cite{TCS98}
\begin{eqnarray}\nno
\mathbb{P}\lp\mX_0 \rightarrow \mX_1 \rp & \leq & \frac{1}{\det{\lp
\mI + (\mX_0- \mX_1)(\mX_0 - \mX_1)^{\dagger}\rp}^{n_r}}.
\end{eqnarray}
The difference codeword pair can be written as:
\begin{eqnarray}%\label{eq:dblastnt2}
\mX_0 - \mX_1 = \Lbr \begin{array}{ccccc}
0 & \cdots & 0 & d^p_{n_t} & d^q_{n_t}  \\
\vdots & \adots & \adots & \adots & 0 \\
0 & d^p_2 & \adots & \adots & \vdots \\
d^p_1 & d^q_1 & 0 &\cdots & 0
\end{array} \Rbr;
\end{eqnarray}
where $\bd^p = [d_1^p,\cdots,d_{n_t}^p]$ and $\bd^q =
[d_1^q,\cdots,d_{n_t}^q]$ are
the codeword difference for a permutation code.\\

\noindent Expanding $(\mX_0- \mX_1)(\mX_0 - \mX_1)^{\dagger}$
in terms of the streams, we get:
\begin{eqnarray}\nno
\det{\lp \mI + (\mX_0- \mX_1)(\mX_0 - \mX_1)^{\dagger}\rp}   \geq
|d_1^pd_2^p\cdots d_{n_t}^p|^2 +|d_1^qd_2^q \cdots d_{n_t}^q|^2.
\end{eqnarray}
The probability of error can be upper
bounded using the union bound:
\begin{eqnarray}\nno
\mathbb{P}_{\mbox{e}} & \leq &
\frac{1}{\SNR^{2\tilde{r}}}\sum_{\mX_0, \mX_1 \neq \mX_0 }
\frac{1}{\lp |d_1^pd_2^p\cdots d_{n_t}^p|^2 +|d_1^qd_2^q\cdots
d_{n_t}^q|^2 \rp^{n_r}}.
\end{eqnarray}
This upper bound can be broken into two summations: one
corresponding to where both the streams are different and the other
summation where one of the streams is the same. Suppose  the same
code is used for both the streams; now the upper bound  can be
simplified:
\begin{eqnarray}\nno
\mathbb{P}_{\mbox{e}} & \leq & \frac{1}{\SNR^{2\tilde{r}}} \sum_{
\bd^p \neq {\bf 0}, \bd^q \neq {\bf 0}} \frac{1}{\lp
|d_1^qd_2^q\cdots d_{n_t}^q|^{2}+|d_1^qd_2^q\cdots
d_{n_t}^q|\rp^{n_r}}\\\nno & & +
\frac{2}{\SNR^{\tilde{r}}}\sum_{\bd^p \neq {\bf
0}}\frac{1}{|d_1^pd_2^p\cdots d_{n_t}^p|^{2n_r}}.
\end{eqnarray}
The arithmetic mean-geometric mean inequality for the term
inside the first summation yields
\begin{eqnarray}\nno
\mathbb{P}_{\mbox{e}} & \dotlq & \frac{1}{\SNR^{2\tilde{r}}} \sum_{
\bd^p \neq {\bf 0}, \bd^q \neq {\bf 0}} \frac{1}{|d_1^qd_2^q\cdots
d_{n_t}^q|^{n_r}|d_1^qd_2^q\cdots d_{n_t}^q|^{n_r} } \\\nno
& & + \frac{2}{\SNR^{\tilde{r}}}\sum_{\bd^p \neq {\bf 0}}\frac{1}{|d_1^pd_2^p\cdots d_{n_t}^p|^{2n_r}}, \\
\nno & = & \lp\frac{1}{\SNR^{\tilde{r}}}\sum_{\bd^p_1 \neq 0}
\frac{1}{|d_1^pd_2^p\cdots d_{n_t}^p|^{n_r}}\rp^2 +
\frac{2}{\SNR^{\tilde{r}}} \sum_{\bd^p_1}\frac{1}{|d_1^pd_2^p\cdots
d_{n_t}^p|^{2n_r}}.
\end{eqnarray}
Now, we use the product distance distribution result in Appendix
\ref{ap:pdd} to separately bound the two summations on the RHS. The
result says that the number of codeword differences pairs with
$|d^p_1\cdots d^p_{n_t}|^2$ less than
$\frac{\SNR^{n_t}}{\SNR^{\tilde{r}-k}}$ is \bqa\nno
\SNR^{\tilde{r}+k}, \eqa for $k$ in $[0,r]$. Using this result,
the first term can be upper bounded as:
\begin{eqnarray}\nno
\lp\frac{1}{\SNR^{\tilde{r}}}\sum_{p_1 \neq 0}
\frac{1}{|d_1^pd_2^p\cdots d_{n_t}^p|^{n_r}}\rp^2 & \dotlq & \lp
\max_{k \in [0,\tilde{r}]} \frac{1}{\SNR^{\tilde{r}}}
\SNR^{k+\tilde{r}}\frac{\SNR^{n_r(\tilde{r}-k)/2}}{\SNR^{n_tn_r/2}}\rp^{2}
\\\nno & = &
\lp\max_{k \in [0,\tilde{r}]}
\SNR^{(1-n_r/2)k}\frac{\SNR^{n_r\tilde{r}/2}}{\SNR^{n_tn_r/2}}
\rp^{2}
\\\nno & = & \SNR^{-(n_r(n_t-\tilde{r}))},
\end{eqnarray}
for $n_r \geq 2$. The second term corresponds to the error when one
of the streams is decoded correctly and can be directly verified  to
be of the correct order. Alternatively,
\begin{eqnarray}\nno
\frac{2}{\SNR^{\tilde{r}}} \sum_{\bd^p_1\neq {\bf
0}}\frac{1}{|d_1^pd_2^p\cdots d_{n_t}^p|^{2n_r}} & \dotlq & \max_{k
\in [0,\tilde{r}]} \frac{1}{\SNR^{\tilde{r}}}\SNR^{k+\tilde{r}}
\frac{\SNR^{n_r(\tilde{r}-k)}}{\SNR^{n_rn_t}},
\\\nno & = &
\max_{k \in [0,\tilde{r}]} \SNR^{k(1-n_r)}
\frac{\SNR^{n_r\tilde{r}}}{\SNR^{n_rn_t}} ,
\\\nno & = & \SNR^{-(n_r(n_t-\tilde{r}))}.
\end{eqnarray}
Thus, combining the two upper bounds, for $n_r \geq 2$ there exists
a code such that the diversity gain is \bqa\nno
n_r(n_t-\tilde{r}).\eqa Taking $\tilde{r}=\frac{n_t+1}{2}r$ proves
Proposition \ref{prop:dblast_firstsegment_rayleigh}.

\subsection{Proof of
Proposition~\ref{prop:timespacerayleigh}}\label{ap:timespace} For
the $(n_t+1) \times 2$ channel, we transposed the code in
\rref{eq:dblastnt2} which was used for achieving the first segment
$n_t \times 2$ channel. The probability of error can be calculated
using a union bound calculation. The pairwise probability of error
is given by \beq \nno 1/\det\lp
\mI+(\mX_0-\mX_1)(\mX_0-\mX_1)^{\dagger}\rp^{n_r}.\eeq Since
\begin{eqnarray}\nno
\det\lp \mI + (\mX_0-\mX_1)(\mX_0-\mX_1)^{\dagger}\rp & = & \det\lp
\mI + (\mX_0-\mX_1)^{\dagger}(\mX_0-\mX_1)\rp,
\end{eqnarray}
the union bound calculation for calculating the probability of error
is exactly the same as same as \rref{eq:dblastnt2} case. Therefore
the diversity obtained by this scheme is given by $n_rn_t -
n_r\tilde{r}$. But in this case we are coding over a block-length of
$n_t$, thus the actual tradeoff curve is $n_rn_t (2-  r)/2$, where
$r\log\SNR$ is the per symbol rate of the channel.

\section{Proof of Proposition \ref{prop:vblast_rayleigh}}\label{ap:qam}
The scheme of sending $n_t$ QAM constellations can be written as
\begin{equation}
\nno \lbr \vq = \sqrt{\frac{\SNR}{\SNR^{\frac{r}{n_t}}}}\lp
i_1,\ldots,i_{2n_t} \rp | i_j \in \zP ~\forall~ j=1,\ldots,2n_t
\rbr,
\end{equation}
where $\zP $ is the integer PAM  constellation with
$\SNR^{\frac{r}{2n_t}}$ points. For a Rayleigh fading channel, the
probability of pairwise error averaged over the fading statistics is
given by \cite{TCS98}:
\begin{equation}\nonumber
\mathbb{P}(\vq_1 \rightarrow \vq_2)   \leq   \Lbr \frac{1}{1 +
||\vq_1 - \vq_2 ||^2}\Rbr^{n_r}.
\end{equation}
Using the union bound the probability of error is bounded by:
\begin{eqnarray*}
\mathbb{P}_e &  \leq & \frac{1}{\SNR^r} \sum_{\vq_1 \not= \vq_2}
\Lbr \frac{1}{1 + ||\vq_1 - \vq_2
||^2}\Rbr^{n_r}, \\
    & \leq & \sum_{\vq \neq {\bf 0}} \frac{1}{||\vq||^{2n_r}},
\end{eqnarray*}
where ${\bf 0}$ is the $2n_t$ dimensional vector of zeros. The
second step follows from the symmetry of the QAM. To compute the
summation in on RHS, we split into a summation over vectors such
that all its components are non-zero and then use the arithmetic
mean-geometric mean (am/gm) inequality. We denote a subset of the
index set, $\{1,2,\cdots,2n_t\}$, by $S$. Then the summation can be
simplified as
\begin{eqnarray*}
\sum_{\vq \neq {\bf 0}} \frac{1}{||\vq||^{2n_r}} & = &
\frac{\SNR^{\frac{rn_r}{n_t}}}{\SNR^{n_r}} \sum_{\lp
i_1,\ldots,i_{2n_t}\rp \not= {\bf 0}}
\frac{1}{\lp|i_1|^2+\cdots+|i_{n_t}|^2\rp^{n_r}}
\\ & = & \frac{\SNR^{\frac{rn_r}{n_t}}}{\SNR^{n_r}}
\sum_{S}\sum_{i_j \not= 0 :  j \in S}
\frac{1}{\lp|i_1|^2+\cdots+|i_{2n_t}|^2\rp^{n_r}}
\\ & \leq & \frac{\SNR^{\frac{rn_r}{n_t}}}{\SNR^{n_r}}
\sum_{S}\sum_{i_j \not= 0 : j \in S} \prod_{j \in
S}\frac{1}{|i_j|^{2n_r/|S|}} \quad{\rm using\ am/gm\ inequality}
\\ & \leq & \frac{\SNR^{\frac{rn_r}{n_t}}}{\SNR^{n_r}}
\sum_{S} \lp \sum_{i_1 \not= 0} \frac{1}{|i_1|^{2n_r/|S|}}\rp^{|S|}
\end{eqnarray*}
Since the range of summation $|i_1|$ is growing with $\SNR$, the
inner summation has different behavior for depending on whether
$|S|$ is larger/smaller than $2n_r$.
\begin{eqnarray*}
\sum_{i_1 \not= 0} \frac{1}{|i_1|^{2n_r/|S|}} & \doteq & \lp
\SNR^{r/2n_t} \rp^{1-2n_r/|S|} \quad {\rm if}~2n_r < |S|
\\ & \doteq & 1 \qquad {\rm otherwise}
\end{eqnarray*}
But because of the definition of $S$, $|S|$ is naturally upper
bounded by $2n_t$. Thus, for $n_r \geq n_t$, the probability of
error can be upper bounded by:
\begin{eqnarray}\label{eq:nrgnt}
\mathbb{P}_e  &  \dotlq & \SNR^{-(n_r-\frac{rn_r}{n_t})}.
\end{eqnarray}
On the other hand, if $n_r <n_t$, then the probability of error can
be upper bounded as:
\begin{eqnarray*}
\mathbb{P}_e  &  \leq & \frac{\SNR^{\frac{rn_r}{n_t}}}{\SNR^{n_r}}
\sum_{S} \lp \sum_{i_1 \not= 0} \frac{1}{|i_1|^{2n_r/|S|}}\rp^{|S|}
\\&\dotlq & \frac{\SNR^{\frac{rn_r}{n_t}}}{\SNR^{n_r}} \sum_{2n_r \leq |S| \leq
2n_t} \lp \SNR^{r/2n_t} \rp^{|S|-2n_r} \\
&  \dotlq & \frac{\SNR^{\frac{rn_r}{n_t}}}{\SNR^{n_r}} \max_{2n_r
\leq |S| \leq 2n_t} \lp \SNR^{r/2n_t} \rp^{|S|-2n_r},
\\ & = & \SNR^{-(n_r-r)}
\end{eqnarray*}

\section{Isotropic MIMO Channels}\label{ap:mimo_isotropic}

We concentrate on the rotationally invariant distributions. For this
class, the singular value distribution determines the channel
statistics completely. Let $f(\vphi)$ be the density function of the
ordered {\em squared} singular values, $\vphi$, of the channel gain
matrix. In terms of notation of Section \ref{sec:mimo}, we have
\bqa\nno \phi_{\ell} & = & \psi_{\ell}^2 \quad {\rm for} \quad
\ell=1,\cdots,\nmin,\eqa where $\psi_\ell$s are the singular values
of $\mH$. In the high $SNR$ regime, we are only interested in the
near zero behavior of $\vphi$. Therefore, in the scaling of
interest, $f$ can be assumed to be of the form:
\begin{eqnarray}\label{eq:generalF}
f(\vphi) & \doteq & \phi_1^{k_1}\cdots\phi_{\nmin}^{k_{\nmin}}
\bone_{\phi_1\leq\phi_2\leq\cdots\leq\phi_{\nmin}}
\end{eqnarray}
This is same as the earlier definition of distribution of the
squared singular values: \beq\nno \prob{\phi_1 \leq
\epsilon_1,\ldots ,\phi_{\nmin}\leq \epsilon_{\nmin}}
\stackrel{.}{=}
 \epsilon_1^{k_1+1}\cdots\epsilon_{\nmin}^{k_{\nmin}+1}, \eeq for
$\eps_1 < \cdots < \eps_{\nmin}$.

For Rayleigh fading distribution, $\vphi$ has the Wishart
distribution which can be reduced to this polynomial form by
ignoring the exponential terms in the Wishart distribution (for the
exact expression, see \cite{ZT03}):
\begin{eqnarray}\label{eq:generalR}
r(\vphi) & \doteq &
\phi_1^{r_{1}}\phi_2^{r_2}\cdots\phi_{\nmin}^{r_{\nmin}}\bone_{\phi_1\leq\phi_2\leq\cdots\leq\phi_{\nmin}}
\end{eqnarray}
where $r_\ell=|n_t-n_r|+2(\ell-1)$.

In this appendix, first we characterize the outage curve in terms of
$k_i$s for general $f$. Then, we use this characterization to
characterize restricted universality for codes based on the V-BLAST
and D-BLAST architecture proposed in Section \ref{sec:vblast} and
Section \ref{sec:dblast} respectively.

\subsection{The outage curve for general fading
distributions}\label{ap:general_outage} For a general fading
distribution, $F$, we want to calculate the probability of outage.
The outage event can be written as: \bqa\nno \sum_{\ell =1}^{\nmin}
\log\lp 1 + \phi_{\ell} \SNR \rp & \leq & r\log\SNR. \eqa If we
write \bqa\label{eq:psitoalpha} \phi_{\ell} = \SNR^{-\alpha_{\ell}},
\eqa then the induced distribution (from \rref{eq:generalF}) on the
ordered vector $\valpha$ is
\begin{eqnarray}
p(\valpha) & \doteq &
\SNR^{-\alpha_1(k_1+1)}\cdots\SNR^{-\alpha_{\nmin}(k_{\nmin}+1)},
\end{eqnarray}
which can obtained by change of variables \rref{eq:psitoalpha}. The
outage probability will be dominated by the $\valpha$ that is on the
boundary of outage and has smallest $\SNR$ exponent. More precisely,
using {\em Laplace's method} as in \cite{ZT03}, the outage curve is
the solution to the optimization problem
\beq\label{eq:alpha_opt}\inf_{\valpha \in A^{'}} \sum_{\ell}
(k_\ell+1)\alpha_{\ell},  \eeq where \bqa\nno A^{'}& = &
\left\{\valpha : \alpha_1 \geq \cdots \geq \alpha_{\nmin} \geq 0
\quad{\rm and}\quad \sum_{\ell}(1-\alpha_\ell)^+ \leq r
\right\}.\eqa The fact that $\alpha_\ell$s are positive uses our
assumption that the singular values have an exponential tail. Let's
assume for some integer $s$, $s \leq r < s+1$. Then, if \bqa
\label{eq:k_is} k_\ell & < & k_{\nmin - s} \quad{\rm for}\quad \ell
= 1,\cdots, \nmin-s-1
\\\nno k_\ell & > & k_{\nmin-s} \quad{\rm for}\quad \ell =
\nmin-s+1,\cdots, \nmin, \eqa then the optimizing $\valpha$ in
\rref{eq:alpha_opt} is given by : \bqa \nno \alpha_{\ell}^* & = & 1
\quad
{\rm for}\quad \ell =1,\cdots, \nmin-s-1 \\
\nno \alpha_{\nmin-s}^* & = & s+ 1 -r   \\\nno \alpha_{\ell}^* & = &
0 \quad {\rm for} \quad \ell = \nmin-s +1 , \cdots, \nmin. \eqa The
corresponding outage curve is given by: \bqa\nno d_{\rm out}(r) & =
& (k_{\nmin-s}+1)(s+ 1 -r) + \sum_{\ell=\nmin-s+1}^{\nmin} \lp
k_\ell + 1 \rp \quad{\rm for} \quad s \leq r < s+1. \eqa

In particular, we would like to stress that if all the $k_i$s are
increasingly ordered then the $\valpha$ that dominates the outage
probability for fading density $f$ is the same one that dominates
the outage probability for for i.i.d.\ Rayleigh fading.

\subsection{Restricted universality of V-BLAST and D-BLAST}

We want to prove that the simple QAM code for the V-BLAST
architecture and codes based on using permutation codes over the
D-BLAST architecture are universal over a class of isotropic fading
distributions. We know that all these codes are tradeoff optimal for
the i.i.d.\ Rayleigh fading channel under the union bound
calculation. We exploit this fact to prove optimality over isotropic
distributions that fade {\em slower} than i.i.d.\ Rayleigh fading.

We denote the diagonal matrices with entries $\vpsi$, the singular
values of the the channel gain matrix, and $\vlambda$, the singular
values of the codeword difference matrix, as $\mPsi$ and $\mLambda$.
Then the probability of pairwise error averaged over the channel
statistics can be written as (see \rref{eq:prwse_pe}):
\begin{eqnarray}
\mathbb{P}_e\lp \mX_A \rightarrow \mX_B \rp & = & \int_{\mH}
Q\lp\sqrt{\frac{\|\mH\mD\|^2}{2}}\rp d\mH
\\\nonumber
& = & \int_{\mPsi}\int_{\mV_1} Q\lp\sqrt{\frac{\|\mPsi\mV_1
\mU_2\mLambda\|^2}{2}}\rp d\mV_1 d\mPsi
\\\nonumber
& = & \int_{\mPhi}\int_{\mV_1} Q\lp\sqrt{\frac{\|\mPhi^{1/2}\mV_1
\mU_2\mLambda\|^2}{2}}\rp d\mV_1 d\mPhi
\\\nonumber
& = & \int_{\mPhi}\int_{\mV} Q\lp{\sqrt{\frac{\|\mPhi^{1/2} \mV
\mLambda\|^2}{2}}}\rp d\mV d\mPhi,
\end{eqnarray}
where the last two steps use the independence of $\mPhi$ and $\mV_1$
and rotational invariance of $\mV_1$ respectively. The integral with
respect to $\mV$ is taken with respect to the Haar measure and does
not depend on the distribution of $\mPhi$ and is only a function of
the realization $\mPsi$ and the code.

Now, the probability of error can be upper bounded using a union
bound \bqa\nno \mathbb{P}_e & \leq & \frac{1}{\SNR^r}
\sum_{\mLambda} \int_{\mPhi}\int_{\mV}
Q\lp{\sqrt{\frac{\|\mPhi^{1/2} \mV \mLambda\|^2}{2}}}\rp d\mV
d\mPsi, \eqa where the summation is over all possible codeword
difference pairs. Since all the terms are positive, interchanging
the order of the summation and integration the union bound can be
written as \bqa\nno \mathbb{P}_e & \leq & \int_{\mPhi} \lp
\frac{1}{\SNR^r} \sum_{\mLambda} \int_{\mV}
Q\lp{\sqrt{\frac{\|\mPhi^{1/2} \mV \mLambda\|^2}{2}}}\rp d\mV \rp
d\mPhi. \eqa

The term  inside the outer integral only depends on the code and the
channel realization $\mPhi$ and not on the fading distribution. We
denote it by $g(\vphi)$. Then the smart union bound can be written
as \bqa\nno\mathbb{P}_e & \leq & \int_{\vphi} g(\vphi) f(\vphi)
d\vphi, \eqa where $f$ is the density function of $\vphi$. Similarly
the upper bound corresponding to the smart union bound is given by
\bqa \label{eq:ssub}  \mathbb{P}_e & \leq & \mathbb{P}\lp{\CH}\rp +
\int_{\mPsi \notin \CH} g(\vphi) f(\vphi) d\vphi, \eqa where $\CH$
is the set of all channel realizations in outage. If we assume that
the union bound is tight for Rayleigh fading, then it implies \bqa
\nno\int_{\vphi \notin \CH} g(\vphi) r(\vphi) d\vphi & \dotlq &
\SNR^{-d_R^*(r)}, \eqa where $r(\vphi)$ is the is density for the
i.i.d.\ Rayleigh fading channel and $d_R^*(r)$ is the corresponding
outage curve. We use $d_F^*(r)$ to denote the optimal curve for a
generic density $f$.

Then, for any $f$ the second term in \rref{eq:ssub} can be upper
bounded as \bqa\nno \int_{\vphi \notin \CH} g(\vphi) f(\vphi) d\vphi
& =  & \int_{\vphi \notin \CH} \frac{f(\vphi)}{r(\vphi)} g(\vphi)
r(\vphi) d\vphi,
\\\nonumber & \leq & \lp\max_{\vphi \notin \CH}\frac{f(\vphi)}{r(\vphi)}\rp
\int_{\vphi \notin \CH} g(\vphi) r(\vphi) d\vphi,
\\\label{eq:upper_bound} & \dotlq & \lp\max_{\vphi \notin
\CH}\frac{f(\vphi)}{r(\vphi)}\rp \SNR^{-d_R^*(r)}. \eqa The
expression to be maximized can be written as (see \rref{eq:generalF}
and \rref{eq:generalR}):
\begin{eqnarray}\label{eq:psi_max2}
\max_{\vphi \notin \CH}\prod_{\ell=1}^{\nmin} \phi_\ell
^{k_\ell-r_\ell} & = & \min_{\vphi \notin \CH}\prod_{\ell=1}^{\nmin}
\phi_\ell ^{u_\ell}.
\end{eqnarray}
where $u_\ell=r_\ell-k_\ell$ Now, we consider the codes from Section
\ref{sec:vblast} and \ref{sec:dblast} and explicitly compute the
maximization \rref{eq:psi_max2}.

\subsubsection*{V-BLAST}
For the last segment of an $n \times n$ channel, none of the
singular values can die completely (\ie, become less than
$\frac{1}{\SNR}$), therefore the no-outage condition can be written
as: \bqa \prod_{\ell=1}^{n} \phi_\ell & \geq
\frac{\SNR^r}{\SNR^n}.\eqa Therefore the minimization
\rref{eq:psi_max2} can be written as
\begin{eqnarray}\label{eq:vblast_opt_phi}
\min_{\prod\limits_{\ell=1}^{n} \phi_\ell \geq
\frac{\SNR^r}{\SNR^n}} \prod_{\ell=1}^{n} \phi_\ell ^{u_\ell},
\end{eqnarray}
with an additional constraint that the $\phi_\ell$s are bounded by
one (using the exponential tail assumption). If we assume that $u_1$
is larger than $u_\ell$ for every $\ell\geq2$, then
\begin{eqnarray}\nno
\min_{\prod\limits_{\ell=1}^{n} \phi_\ell \geq
\frac{\SNR^r}{\SNR^n}} \prod_{\ell=1}^{n} \phi_\ell ^{u_\ell} & \leq
& \frac{\min\limits_{\prod\limits_{\ell=1}^{n} \phi_\ell \geq
\frac{\SNR^r}{\SNR^n}}\lp\prod\limits_{\ell=1}^{n}
\phi_\ell\rp^{u_1}}{\max\limits_{\vphi}\prod\limits_{\ell=2}^{n}
\phi_\ell^{u_1-u_\ell}} ,
\end{eqnarray}

If we assume that $u_1 \geq 0$ and $u_1-u_\ell \geq 0$ for every
$\ell\geq2$, then the optimizing solution is given by:
\begin{eqnarray}\nno \phi_1^* & = & \frac{\SNR^r}{\SNR^n} \\\nno
\phi_{\ell}^* & = & 1 \quad {\rm for}\quad \ell
=2,\cdots,L.\end{eqnarray} This optimal point is same as the point
(in terms of $\valpha$), that optimized the outage probability
calculation in \rref{eq:alpha_opt}. Then, at the optimal point we
can write:
\begin{eqnarray}\nno
\prod_{\ell=1}^{n} \phi_{\ell}^{*^{k_\ell -r_\ell}}&  = &
\frac{\prod_{\ell=1}^{n}
\phi_{\ell}^{*^{k_\ell+1}}}{\prod_{\ell=1}^{n}
\phi_{\ell}^{*^{r_\ell+1}}}
\\\nno & =& \frac{\SNR^{-d_F^*(r)}}{\SNR^{d_R^*(r)}}.
\end{eqnarray}
Therefore, using \rref{eq:upper_bound} and \rref{eq:ssub} the
probability of error can be upper bounded by \bqa\nno \mathbb{P}_e &
\dotlq & \SNR^{-d_F^*(r)} + \frac{\SNR^{-d_F^*(r)}}{\SNR^{d_R^*(r)}}
\SNR^{-d_R^*(r)}
\\\nno & \dotlq & \SNR^{-d_F^*(r)}.
\eqa Thus, the code is also tradeoff optimal for the channel with
fading density $f$, where $f$ satisfies the following conditions:
\begin{eqnarray*} r_1 & \geq & k_1 \\\nno
r_1 -k_1 & \geq & r_1 +2(\ell-1) - k_\ell \quad{\rm for}\quad \ell
=2,\cdots,L.
\end{eqnarray*}
Combining these two conditions, we get
\begin{eqnarray*}
 k_\ell  - 2(\ell-1) & \geq &  k_1  \quad{\rm for}\quad \ell
=2,\cdots,L.\\\nno k_1 & \leq & 0.
\end{eqnarray*}

\subsubsection*{D-BLAST}
For the last segment of an $n_t \times 2$ channel, none of the
singular values can fade completely (\ie become less than
$\frac{1}{\SNR}$), and hence the no-outage condition can be written
as: \bqa  \phi_1 \phi_2 & \geq \frac{\SNR^r}{\SNR^2}.\eqa This means
that
the minimization \rref{eq:psi_max2} can be written as
\begin{eqnarray}
\min_{\phi_1 \phi_2 \geq \frac{\SNR^r}{\SNR^2}} \phi_1^{u_1}\phi_2
^{u_2},
\end{eqnarray}
Now, this optimization problem is the same as the V-BLAST optimization
problem in
\rref{eq:vblast_opt_phi}, with $n=2$. Hence, the optimality condition
on $k_1$ and $k_2$ turns out to be
\begin{eqnarray*}
 k_2 - k_1 & \geq & 2\\\nno k_1 & \leq & 0
\end{eqnarray*}

\bibliographystyle{IEEEtran}
\bibliography{myref}

\end{document}